\documentclass{pasj01}
\usepackage[T1]{fontenc}
\Received{$\langle$reception date$\rangle$}
\Accepted{$\langle$acception date$\rangle$}
\Published{$\langle$publication date$\rangle$}
\begin{document}

\title{The Quasar Luminosity Function at Redshift 4 \\ with Hyper Suprime-Cam Wide Survey}
\author{Masayuki \textsc{Akiyama}\altaffilmark{1},
Wanqiu \textsc{He}\altaffilmark{1},
Hiroyuki \textsc{Ikeda}\altaffilmark{2},
Mana \textsc{Niida}\altaffilmark{3},
Tohru \textsc{Nagao}\altaffilmark{4},
James \textsc{Bosch}\altaffilmark{5},
Jean \textsc{Coupon}\altaffilmark{6},
Motohiro \textsc{Enoki}\altaffilmark{7},
Masatoshi \textsc{Imanishi}\altaffilmark{2}, 
Nobunari \textsc{Kashikawa}\altaffilmark{2},
Toshihiro \textsc{Kawaguchi}\altaffilmark{8},
Yutaka \textsc{Komiyama}\altaffilmark{2,9},
Chien-Hsiu \textsc{Lee}\altaffilmark{10},
Yoshiki \textsc{Matsuoka}\altaffilmark{4,2},
Satoshi \textsc{Miyazaki}\altaffilmark{2,9},
Atsushi J. \textsc{Nishizawa}\altaffilmark{11},
Masamune \textsc{Oguri}\altaffilmark{12,13,14},
Yoshiaki \textsc{Ono}\altaffilmark{15},
Masafusa \textsc{Onoue}\altaffilmark{9,2},
Masami \textsc{Ouchi}\altaffilmark{15,14},
Andreas \textsc{Schulze}\altaffilmark{2},
John D. \textsc{Silverman}\altaffilmark{14},
Manobu M. \textsc{Tanaka}\altaffilmark{16,17},
Masayuki \textsc{Tanaka}\altaffilmark{2},
Yuichi \textsc{Terashima}\altaffilmark{3},
Yoshiki \textsc{Toba}\altaffilmark{18},
Yoshihiro \textsc{Ueda}\altaffilmark{19}
}
\altaffiltext{1}{Astronomical Institute, Tohoku University, Aramaki, Aoba-ku, Sendai, 980-8578}
\altaffiltext{2}{National Astronomical Observatory of Japan, 2-21-1, Osawa, Mitaka, Tokyo 181-8588}
\altaffiltext{3}{Department of Physics, Ehime University, Bunkyo-cho, 2-5, Matsuyama, Ehime 790-8577}
\altaffiltext{4}{Research Center for Space and Cosmic Evolution, Ehime University, Bunkyo-cho 2-5, Matsuyama, 790-8577}
\altaffiltext{5}{Department of Astrophysical Sciences, Princeton University, 4 Ivy Lane, Princeton, NJ 08544, USA}
\altaffiltext{6}{Department of Astronomy, University of Geneva, ch. d\'Ecogia 16, 1290 Versoix, Switzerland}
\altaffiltext{7}{Faculty of Business Administration, Tokyo Keizai University, Kokubunji, Tokyo, 185-8502}
\altaffiltext{8}{Department of Economics, Management and Information Science, Onomichi City University, Hisayamada 1600-2, Onomichi, Hiroshima 722-8506}
\altaffiltext{9}{Department of Astronomical Science, Graduate University for Advanced Studies (SOKENDAI), 2-21-1, Osawa, Mitaka, Tokyo 181-8588}
\altaffiltext{10}{Subaru Telescope, National Astronomical Observatory of Japan, 650 N Aohoku Pl, Hilo, HI 96720, USA}
\altaffiltext{11}{Institute for Advanced Research, Nagoya University, Furocho Chikusa-ku, Nagoya, 464-8602}
\altaffiltext{12}{Research Center for the Early Universe, The University of Tokyo, Tokyo 113-0033}
\altaffiltext{13}{Department of Physics, The University of Tokyo, Tokyo 113-0033}
\altaffiltext{14}{Kavli Institute for the Physics and Mathematics of the Universe (Kavli IPMU, WPI), The University of Tokyo, 5-1-5 Kashiwanoha, Kashiwa, Chiba 277-8583}
\altaffiltext{15}{Institute for Cosmic Ray Research, The University of Tokyo, 5-1-5, Kashiwanoha, Kashiwa, Chiba 277-8582}
\altaffiltext{16}{High Energy Accelerator Research Organization, Oho 1-1, Tsukuba, Ibaraki, 305-0801}
\altaffiltext{17}{The Graduate University for Advanced Studies, Oho 1-1, Tsukuba, Ibaraki, 305-0801}
\altaffiltext{18}{Academia Sinica Institute of Astronomy and Astrophysics, P.O. Box 23-141, Taipei 10617, Taiwan}
\altaffiltext{19}{Department of Astronomy, Kyoto University, Kitashirakawa-Oiwake-cho, Sakyo-ku, Kyoto, 606-8502}

\email{akiyama@astr.tohoku.ac.jp}


\maketitle

\begin{abstract}
We present the luminosity function of $z=4$ quasars based on the
Hyper Suprime-Cam Subaru Strategic Program Wide layer
imaging data in the $g$, $r$, $i$, $z$, and $y$ bands
covering 339.8 deg$^{2}$. From stellar objects, 1666 $z\sim4$ quasar 
candidates are selected by the $g$-dropout selection down to $i=24.0$ mag. 
Their photometric
redshifts cover the redshift range between 3.6 and 4.3 with an average
of 3.9. In combination with the quasar sample from the Sloan Digital
Sky Survey in the same redshift range, the quasar luminosity function
covering the wide luminosity range of $M_{\rm 1450}=-22$ to $-29$ mag
is constructed. It is well described by
a double power-law model with a knee at $M_{\rm 1450}=-25.36\pm0.13$ mag
and a flat faint-end slope with a power-law index of $-1.30\pm0.05$. 
The knee and faint-end slope show no clear evidence of redshift
evolution from those at $z\sim2$. The flat slope implies that the UV luminosity density
of the quasar population is dominated by the quasars around the knee, 
and does not support the steeper faint-end slope at higher redshifts reported at $z>5$.
If we convert the $M_{\rm 1450}$ luminosity function
to the hard X-ray 2--10~keV luminosity function using the
relation between UV and X-ray luminosity of quasars and its scatter,
the number density of UV-selected quasars matches well with 
that of the X-ray-selected AGNs above the knee of the luminosity
function. Below the knee, the UV-selected quasars show a 
deficiency compared to the hard X-ray luminosity function. The deficiency
can be explained by the lack of obscured AGNs among the UV-selected quasars.
\end{abstract}

\section{Introduction}

After the discovery that every massive galaxy harbors a
super massive black hole (SMBH) in its center,
the issue of how these SMBHs formed and evolved over cosmic
history became one of the major unanswered questions in observational
cosmology. The cosmological evolution of the AGN luminosity function, 
which reflects the growth history of SMBHs through accretion, has been intensively
investigated using large AGN samples (e.g, 
\cite{Richards2006}; \cite{Croom2009}; \cite{McGreer2013}; \cite{Ross2013};
\cite{Ueda2014}). 
There is an overall trend that the number density of more luminous AGNs
peaks at higher redshift, so-called "down-sizing" and "antihierarchical" growth, 
and the number density of the most luminous AGNs, i.e. quasars, shows rapid
decline from the peak redshift, $z\sim3$ to the local universe. 

On the other hand, the number
density of less-luminous AGNs beyond its peak redshift would show a hint
of milder decline \citep{Ueda2014}. Such evolution would be the first evidence of a
different evolutionary trend in the early universe that less-luminous
AGNs are more numerous at first, then luminous AGNs grow later, 
i.e. "up-sizing" and "hierarchical" growth of SMBHs. In order to disclose
the discussion of the evolutionary trend seen in the early universe, it is
important to determine the shape of the AGN luminosity function, especially around its knee, 
at each redshift, and examine the redshift evolution of the shape.

The faint-end slope of the AGN luminosity function in the early universe is also important
to evaluate the contribution of AGNs to the UV ionizing photon budget.
Recently utilizing X-ray-selected AGNs, \citet{Giallongo2015} derived $z=4,5,6$
AGN luminosity functions with a steep slope and high number density in the faint-end. If we take
the high number density of less-luminous AGNs at face value, the AGN
emissivity of UV ionizing photons would be as high as the value required
to keep the intergalactic medium highly ionized, and can significantly 
contribute to the cosmic reionization.

The strategic survey program of the Subaru telescope with the Hyper Suprime-Cam
(HSC-SSP; \cite{Aihara2017a}), which started in Mar. 2014 and is assigned 300 nights for 5 years,
provides us a unique opportunity to determine the shape of the quasar luminosity function 
in the early universe with unprecedented accuracy. An overview of the camera and details
of the dewar system are given in \citet{Miyazaki2017} and \citet{Komiyama2017}, respectively.
The Wide-layer component
of the survey covers 1,400 deg$^{2}$ in the equatorial region 
down to 26.8, 26.4, 26.4, 25.5, and 24.7 mags in the $g$,$r$,$i$,$z$,$y$ bands,
respectively, with 5$\sigma$ for point sources in the 5 year survey. 
The latest internal release of the data, S16A-Wide2, covers 339.8 deg$^{2}$,
including the edge regions where the final depth have not been achieved \citep{Aihara2017b}.
The depths of the Wide-layer component are about 1 mag deeper than previous 
wide-field surveys covering similar area 
(e.g. Canada-France-Hawaii Telescope Legacy Survey wide fields
covering 150 deg$^{2}$).
Furthermore, the image quality is better than other wide-field surveys;
the median seeing size in the $i$ band is \timeform{0.61"} \citep{Aihara2017b}.
Thanks to the depth and image quality of the data, we can select 
candidates of $z=4$ quasars with stellarity and $g$-dropout selections
reliably down to $i=24.0$ mag, which corresponds to $M_{\rm 1450}=-22$ mag, i.e. 
well below the knee of the quasar luminosity function.
In combination with the spectroscopically identified $z=4$ quasars
from the Sloan Digital Sky Survey (SDSS) data release 7 (DR7) \citep{Schneider2010}, we can construct
the $z=4$ quasar luminosity function covering a wide-luminosity range.

In this paper, we select candidates of $z\sim4$ quasars applying a $g$-band
dropout selection to objects with stellar morphology. The selection
criteria and definition of the sample are described in section 2. We evaluate
the effective survey area by constructing $z\sim4$ quasar photometric models based on 
a SED library of quasars and a noise model of the HSC Wide-layer dataset.
The statistical contamination rate of the sample is estimated with photometric
data of Galactic stars and galaxies. The effective survey area and 
the contamination rate are discussed in section 3. Because most of the 
selected candidates do not have spectroscopic information, we derive their
photometric redshifts with a Bayesian method using the
library of quasar photometric models. In order to construct the $z=4$ quasar
luminosity function covering a wide luminosity range, we combine our results with a sample
of luminous quasars in the same redshift range utilizing the SDSS DR7 quasar catalog.
The properties of the sample
and the derivation of the $z=4$ quasar luminosity function is described in section 4.
In section 5, we discuss the evolution of the quasar luminosity function 
at $z>2$, and we compare the $z=4$ quasar luminosity function with the
luminosity function of X-ray selected AGNs at $z=4$.
Throughout the paper, we use cosmological parameters of
$H_{0}=70$ km s$^{-1}$ Mpc$^{-1}$, $\Omega_{m}=0.3$, and
$\Omega_{\lambda}=0.7$. All magnitudes are described in the AB magnitude
system.

\section{Sample definition}

\subsection{HSC-SSP Wide-layer catalog}
\label{sec:database}

Candidates of $z\sim4$ quasars have been selected from
the Wide-layer component of the HSC-SSP database.
The S16A-Wide2 internal release of the survey 
covers an area of 339.8 deg$^{2}$, where the $g$, $r$, $i$,
$z$, and, $y$ bands data are available. The area
includes edge regions where the integration time 
has not reached the target value and the final
depth has not been achieved, therefore the area is
larger than the full-color and full-depth area.
The data are reduced with hscPipe-4.0.2 \citep{Bosch2017}.

In the S16A-Wide2 release, the Wide-layer survey
data is separated into 7 continuous fields. Rough
central coordinates of the fields are summarized
in table~\ref{tab:subregions}. We divide
the fields into 4 sub-regions of WideA-D.
The total area of each sub-region is listed in table~\ref{tab:subregions}. 
The total area 
only considers area covered by all 5 bands.
We remove some patches where the color sequence
of stars brighter than $i_{\rm PSF}<22$ mag 
shows significant offset from that 
expected from the Gunn-Stryker stellar
spectro-photometric library \citep{Gunn1983};
patches which have offset in the {\bf patch\_qa}
table larger than 0.075 mag either in the
$g-r$ vs. $r-i$, $r-i$ vs. $i-z$, or $i-z$ vs. $z-y$
color-color plane (see section 5.8.4 in \citet{Aihara2017b}).
We also remove tract 8284, because the photometric data in the
region are unreliable.

We first select objects which are not flagged
with saturation, bad pixel, and cosmic-ray
hit and located at the edge of the detector.
In the HSC pipeline, crowded objects are
deblended by a deblending process. We consider
the objects after the deblending process.
The selection conditions are 
summarized in table~\ref{tab:TIDYselection}.

We use PSF magnitudes for stellar objects.
The PSF magnitude is determined by fitting a model PSF
at each position to an image of an object. For discussions on extended
objects, we refer their CMODEL magnitudes.
CMODEL magnitudes are determined by at first
fitting exponential and de Vaucouleurs profiles separately, 
then fitting them together \citep{Bosch2017}. Both profiles are 
convolved with the model PSF at the position of each object.
The profiles of stellar objects are fitted with 
exponential and/or de Vaucouleurs profiles with radius of 0, 
CMODEL magnitudes should be consistent with the PSF magnitudes.
However, PSF magnitudes are 
more stable, because the fitting is not affected by the 
uncertainty in the profile model.
All the magnitudes are corrected for Galactic extinction
based on dust maps by \citet{Schlegel1998}.

\begin{table*}
  \caption{Sub-regions defined in this paper}\label{tab:subregions}
  \begin{center}
    \begin{tabular}{clccrrrrr}
\hline
	    Sub-regions & \multicolumn{1}{c}{Field name}   & Central coordinate                 & Galactic coordinate              & \multicolumn{1}{c}{Area}        & \multicolumn{1}{c}{Eff.Area\footnotemark[$\dag$]} & \multicolumn{1}{c}{Eff.Area\footnotemark[$\ddag$]} & \multicolumn{1}{c}{$N_{\rm org}$} & \multicolumn{1}{c}{$N_{\rm sample}$} \\
	     &          &  (RA,Dec)                          &  ($l$, $b$)                      & \multicolumn{1}{c}{(deg$^{2}$)} & \multicolumn{1}{c}{(deg$^{2}$)} & \multicolumn{1}{c}{(deg$^{2}$)} &               &                  \\
\hline
 WideA       & XMM-LSS  &  (\timeform{34},\timeform{-4})   & (\timeform{168},\timeform{-59}) &  62.8       & 41.6     & 35.1     &       594     &           372    \\
 WideB       & GAMA09H  &  (\timeform{135},\timeform{0})  & (\timeform{229},\timeform{28})  &  81.1       & 45.2     & 38.1     &       747     &           341    \\
 WideC       & WIDE12H  &  (\timeform{180},\timeform{0})  & (\timeform{276},\timeform{60})  &  107.9      & 68.5     & 58.6     &       984     &           527    \\
             & GAMA15H  &  (\timeform{218},\timeform{0})  & (\timeform{349},\timeform{54})  &             &          &          &               &                  \\
 WideD       & HECTOMAP &  (\timeform{243},\timeform{43}) & (\timeform{68},\timeform{47})   &  88.0       & 46.4     & 40.2     &       902     &           428    \\
             & VVDS     &  (\timeform{340},\timeform{0})  & (\timeform{68},\timeform{-48})  &             &          &          &               &                  \\
             & AEGIS    &  (\timeform{215},\timeform{52}) & (\timeform{95},\timeform{60})   &             &          &          &               &                  \\
\hline
 Total       &          &                                   &                                   & 339.8       & 201.7    & 172.0    &      3,227    &         1,668\footnotemark[$*$]    \\
\hline
    \end{tabular}
  \end{center}
\begin{tabnote}
   \footnotemark[$\dag$] Effective survey area without masks around $i<22$ mag objects.
   \footnotemark[$\ddag$] Effective survey area with masking around $i<22$ mag objects.
   \footnotemark[$*$] Including 2 spectroscopically identified quasars at $z\sim1$, see section 4.1.
\end{tabnote}
\end{table*}

\begin{table}
  \caption{Database selection criteria}\label{tab:TIDYselection}
  \begin{center}
    \begin{tabular}{ll}
\hline
 \multicolumn{1}{c}{Flag}         & \multicolumn{1}{c}{condition} \\
\hline
 detect\_is\_primary              & True     \\
 flags\_pixel\_edge               & not True \\
 flags\_pixel\_saturated\_center  & not True \\
 flags\_pixel\_cr\_center         & not True \\
 flags\_pixel\_bad                & not True \\
% flags\_pixel\_bright\_object\_center & not True \\
\hline
    \end{tabular}
  \end{center}
\end{table}

\subsection{Selecting stellar objects}
\label{sec:SGclass}

We consider AGNs whose morphology is dominated
by their nuclear stellar component as quasars. 
Stellar objects are selected by comparing the
second order adaptive moments of an object with those
of the PSF at the position of the object. 
We employ the adaptive moments measured in the $i$ band,
because $i$ band images of the Wide-layer survey 
are selectively taken under good seeing conditions
for weak-gravitational lensing study. The median
seeing size of the $i$ band images is \timeform{0.61"}.
The adaptive moment evaluated with the
algorithm described in \citet{Hirata2003} is available
in the HSC-SSP database. We use the conditions,

\begin{eqnarray}
{\rm ishape\_hsm\_moment\_11} &/& {\rm ishape\_hsm\_psfmoment\_11} \nonumber \\
	                                                         &<& 1.1 \\ 
{\rm ishape\_hsm\_moment\_22} &/& {\rm ishape\_hsm\_psfmoment\_22} \nonumber \\
	                                                         &<& 1.1 
\end{eqnarray}

to select stellar objects. We remove objects
whose adaptive moments are not measured correctly and 
listed as "nan".
We do not refer CLASSIFICATION\_EXTENDEDNESS,
which is provided in the database,
because the classification criterion is more optimized
to select extended objects, and contamination of extended
objects among stellar object is not negligible 
compared to the selection criteria applied in this paper.

The incompleteness and contamination of the above criteria are
evaluated with the simulated images of the Wide-layer dataset 
in the Cosmic Evolution Survey (COSMOS) region, where $i$-band imaging data with 
Advanced Camera for Surveys (ACS)
on Hubble Space Telescope (HST) is available. The simulated images
are constructed with selected images
from ultra-deep HSC-SSP survey in the COSMOS region.
Three stacked images simulating good, median, and bad seeing
conditions during the Wide-layer observations are provided in the database
as the COSMOS wide-depth stacks, they have FWHMs of
\timeform{0.5"}, \timeform{0.7"}, and \timeform{1.0"}. It should be
noted that the $i$ band images are selectively taken under good
seeing conditions, therefore most of the data are taken under
the good and median seeing conditions, and rarely taken under
the bad seeing condition (see figure 3 of \cite{Aihara2017b}).
The ACS image is deeper than 
the Wide-layer survey, and stellarity classification is
available in the object catalog \citep{Leauthaud2007}. The classification
is more robust than that based on the HSC images thanks to the
sharper PSF with HST. 
The incompleteness is defined by the fraction of ACS 
stellar objects mis-classified as extended objects 
with the above criteria on the HSC catalog. 
The contamination is defined by the fraction of the ACS
extended objects that are classified as stellar with the
above criteria among the entire HSC stellar objects.

\begin{figure*}
 \begin{center}
  \includegraphics[scale=0.75, angle=-90]{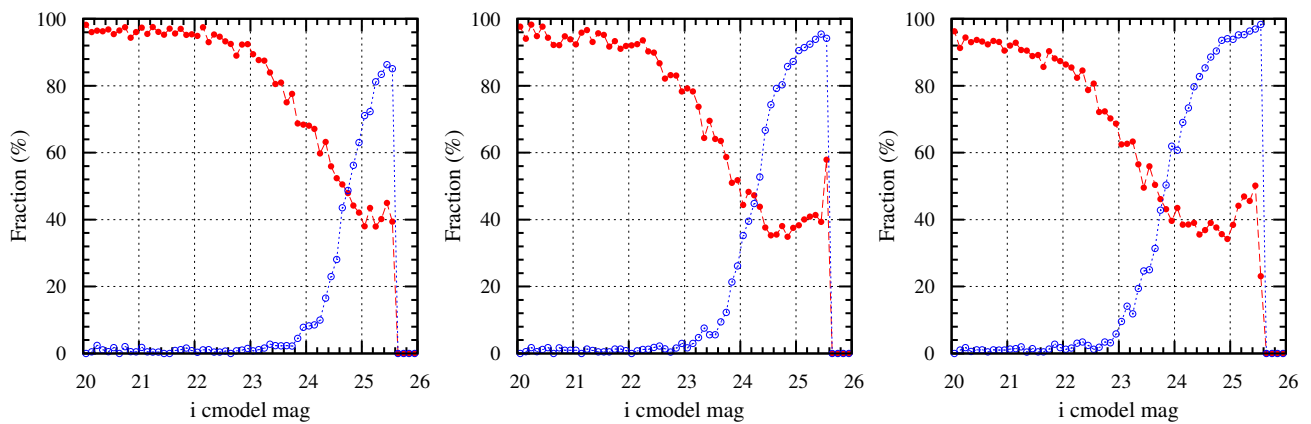}
 \end{center}
\caption{
Fractions of the ACS stellar objects classified as stellar in the HSC
stellar selection (red filled circles connected with dashed line), and of the 
ACS extended objects contaminating among the HSC stellar objects
as a function of $i$-band magnitude (blue open circles connected with dotted line). Left) based on the
stack simulating the best seeing conditions with FWHM=\timeform{0.5"},
middle) simulating the median seeing conditions with FWHM=\timeform{0.7"}, and
right) the bad seeing conditions with FWHM=\timeform{1.0"}. It should be noted
that $i$ band data in the Wide-layer are mostly taken under the good and median
seeing conditions.
\label{fig:HSCUDEEP0_STAR11_ACSstellar_frac}}
\end{figure*}

Figure~\ref{fig:HSCUDEEP0_STAR11_ACSstellar_frac} shows
the resulting incompleteness and contamination of the
classification. In the 
median condition, the completeness is more than 80\%
down to $i=23$ mag and decreases to 40\% at $i=24$ mag, 
while the fraction of contaminating objects
is less than 5\% down to $i=23$ mag, and increases
to 30\% at $i=24$ mag. In the good and bad seeing 
conditions, the completeness and contamination vary 
accordingly. As we explained above, most of the $i$ band
data are taken under the condition simulated as the good
and median conditions, therefore hereafter we refer the
completeness and contamination evaluated with the median
condition. We need to note that the current survey area includes
regions with shallower depth, and in such regions, 
the above incompleteness and contamination rate may not be applicable.
Because the contamination rate increases rapidly at
$i>24$ mag, hereafter we only consider objects brighter than $i=24$ mag.

\subsection{Color selection criteria for $z\sim4$ quasars}
\label{sec:colorsel}

\begin{figure*}
 \begin{center}
  \includegraphics[scale=0.6]{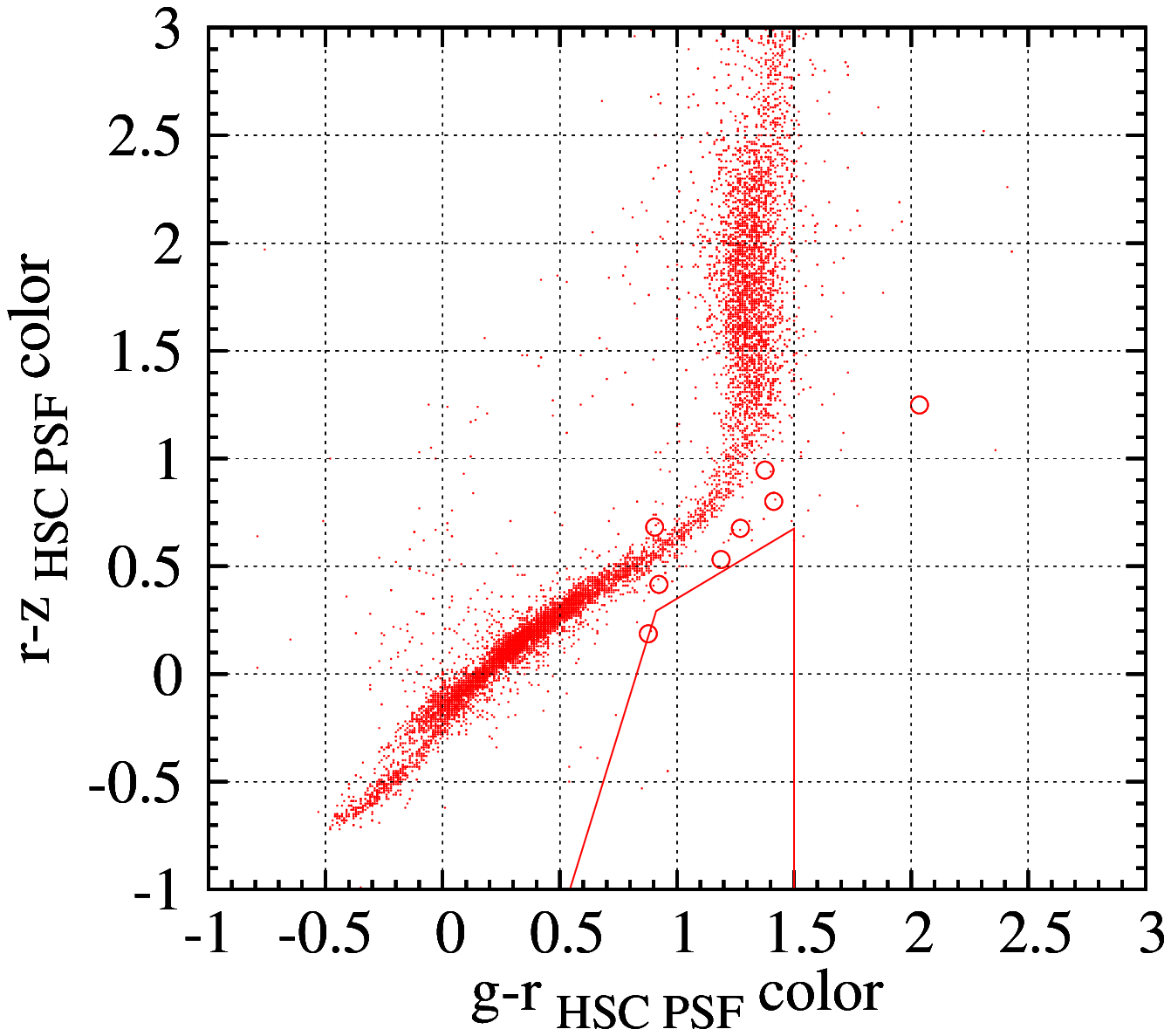}
  \includegraphics[scale=0.6]{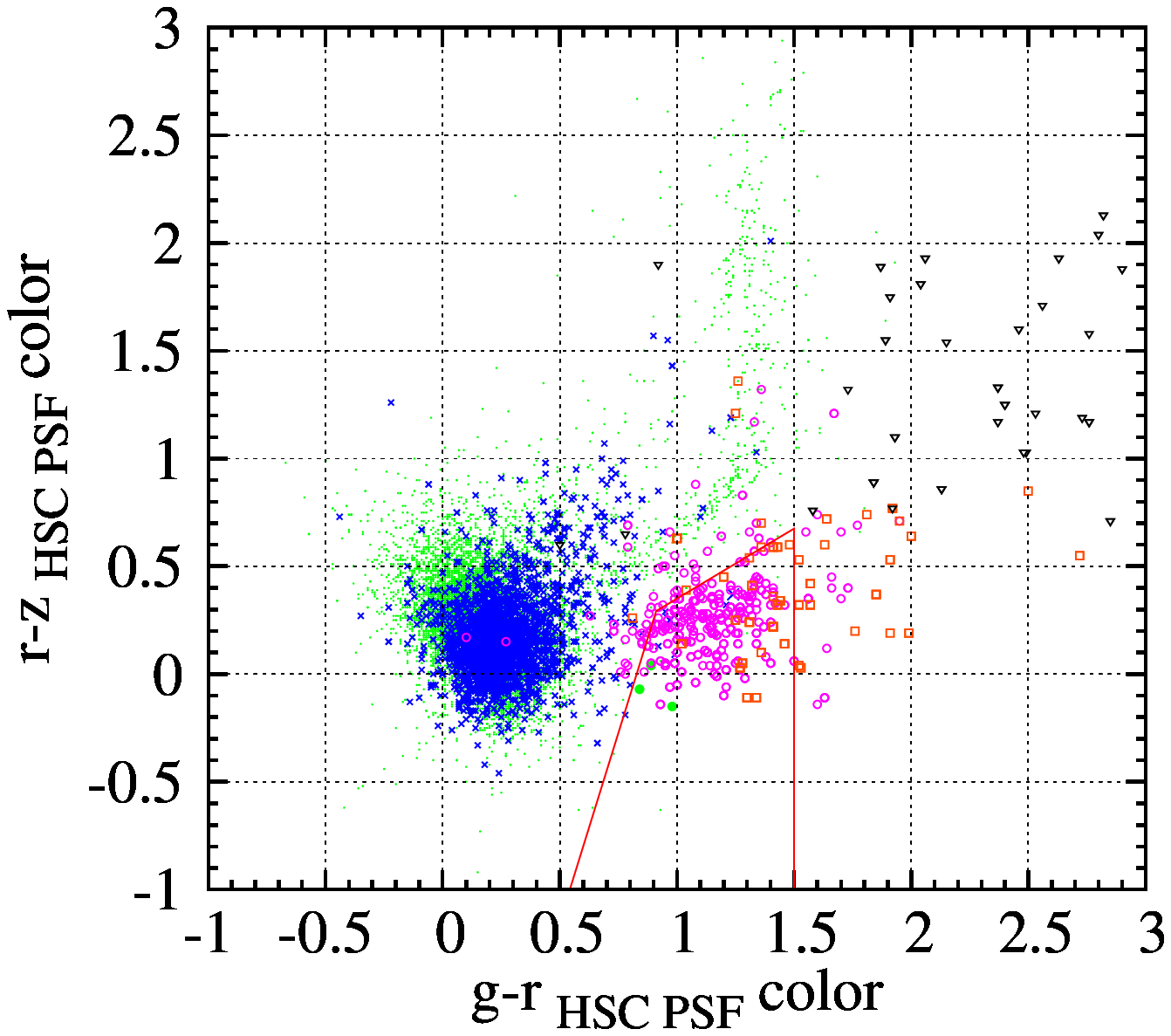}
 \end{center}
\caption{
Distribution of HSC stellar objects with spectroscopic
information on $g-r$ vs. $r-z$ color-color diagram.
Left) distributions of Galactic stars.
Red dots: Galactic stars,
red large open circles
represent HSC colors of Galactic stars identified in 
the $z\sim4$ quasar survey in the COSMOS region \citep{Ikeda2011}.
Right) distributions of extragalacitc objects.
Green dots: objects at $0<z<2.5$, blue crosses: objects at $2.5<z<3.5$, 
pink open circles: objects at $3.5<z<4.0$, orange open squares: objects at $4.0<z<4.5$, 
and black triangles: objects at $4.5<z$. 
Red solid lines in both of the panels indicate the color 
selection criteria used in this paper. 
\label{fig:ZSPstellar_HSCWIDE_gr_rz}}
\end{figure*}

Candidates of $z\sim4$ quasars are selected from
stellar objects on the $g-r$ vs. $r-z$ color-color diagram.
Figure~\ref{fig:ZSPstellar_HSCWIDE_gr_rz} summarizes
the distribution of the S16A-Wide2 stellar objects with known spectroscopic
information on the color-color diagram. The selection criteria
are determined aimed at including as many quasars between
$3.5<z<4.0$ as possible, while minimizing the contamination 
by other objects. 
The determined selection criteria are shown with solid
lines. They are
\begin{eqnarray}
0.65 (g-r) - 0.30 &>& (r-z) \\
3.50 (g-r) - 2.90 &>& (r-z) \\
(g-r) &<& 1.50. 
\end{eqnarray}

We use the third selection criterion to limit the redshift
range of the sample to $z\sim4.5$.

In order to constrain the quasar luminosity function at $z\sim4$
without conducting further spectroscopic follow-up observation, we set
relatively tight color selection criteria to minimize
contamination of red Galactic stars. For example, in \citet{Ikeda2011},
they use wider selection criteria to select $z\sim4$ quasar candidates
for spectroscopic follow-up observations in the COSMOS region, 
and they found 
a significant number of contamination by Galactic stars. 
The HSC colors of their Galactic stars are shown with open red circles
in the left panel of the figure. We set the tighter selection criteria not to include
these stars by loosing a non-negligible fraction of $z\sim4$ quasars
with known spectroscopic redshift. The fraction of quasars
missed in the color selection is accounted for in the statistical 
evaluation of the survey effective area discussed in the next section.

\begin{figure}
 \begin{center}
  \includegraphics[scale=0.6]{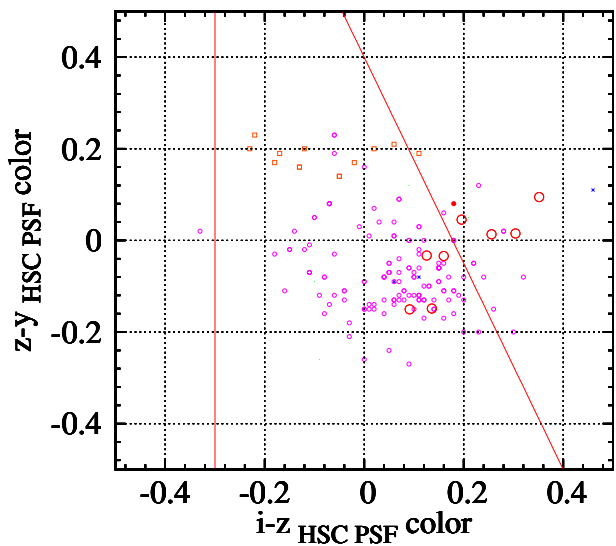}
 \end{center}
\caption{
Distribution of HSC stellar objects with spectroscopic
information on $i-z$ vs. $z-y$ color-color diagram.
Symbols are the same as figure~\ref{fig:ZSPstellar_HSCWIDE_gr_rz}.
Only objects that meet the $g-r$ vs. $r-z$ color
selection criteria are shown, except for 
Galactic stars identified in the
$z\sim4$ quasar survey in the COSMOS region \citep{Ikeda2011}.
Red solid lines indicate the color selection criteria used in this paper.
\label{fig:ZSPstellar_HSCWIDE_iz_zy}}
\end{figure}

In addition to the criteria on the $g-r$ vs $r-z$ plane, 
we also apply criteria on the $i-z$ vs $z-y$ plane with
\begin{eqnarray}
-2.25 (i-z) + 0.400 &>& (z-y) \\
(i-z) &>& -0.3. 
\end{eqnarray}
These criteria are necessary to further remove contamination by
red Galactic stars, and to remove some outliers with unreliable 
photometry. The distribution of the spectroscopically identified stellar objects that meet the
$g-r$ vs. $r-z$ color selection criteria in the $i-z$ vs. $z-y$ plane 
is shown in figure~\ref{fig:ZSPstellar_HSCWIDE_iz_zy}. The color selection
in the $i-z$ vs. $z-y$ plane removes reddened quasars. 
In addition to the above color selection criteria, in order not to be affected by objects
with low signal-to-noise ratio, we only consider objects with magnitude error less
than 0.1 mag in both of the $r$ and $i$ bands.

\begin{figure*}
 \begin{center}
  \includegraphics[scale=0.60]{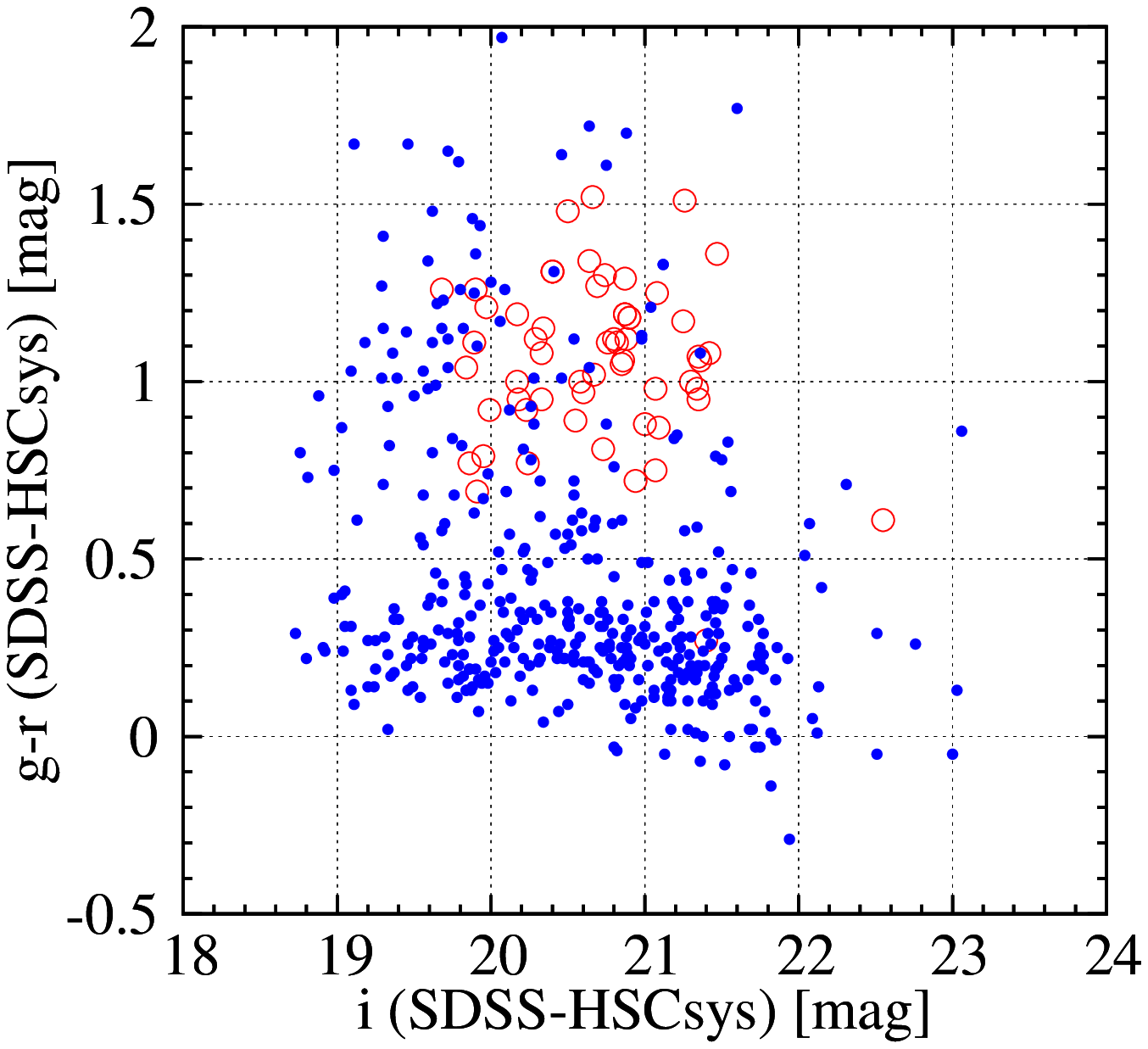}
  \includegraphics[scale=0.60]{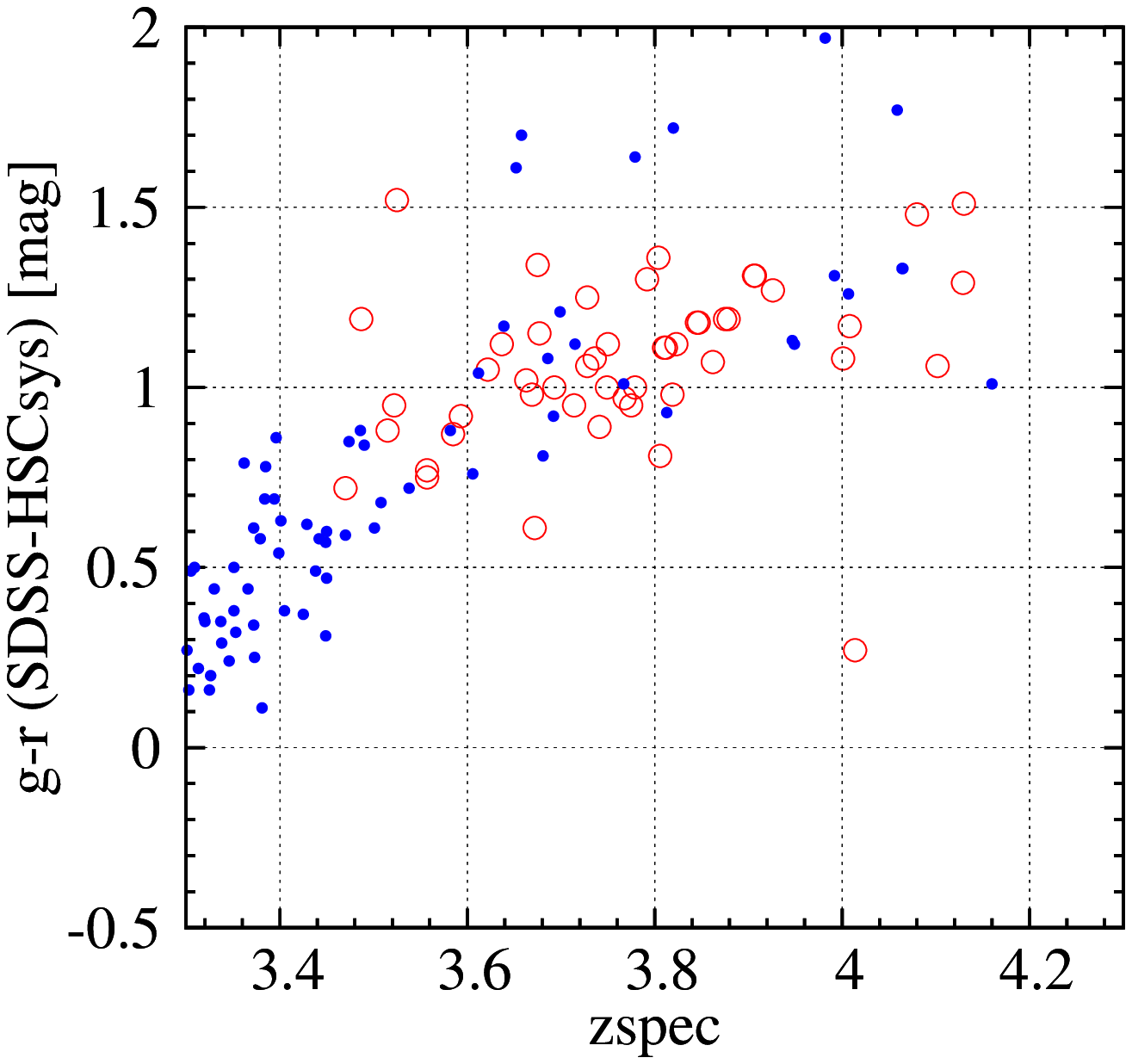}
 \end{center}
\caption{
Left) $i$ band magnitude vs. $g-r$ color of the 
SDSS quasars at $z=3.0-4.5$ in the S15B-Wide2 coverage.
Red open and blue filled circles represent
SDSS quasars selected and missed
by our HSC $z=4$ quasar selection criteria, respectively.
Objects brighter than $i<20.0$ mag are not selected
by the HSC selection due to the effect of the saturation.
Right) $z_{\rm spec}$ vs. $g-r$ color of the SDSS
quasars fainter than $i>20.0$ mag. Red open
and blue filled circles are the same as the left panel.
$g-r$ colors in the HSC systems are derived from 
the SDSS photometry.
\label{fig:SDSS_HSCz4_zsp_gr}}
\end{figure*}

In the left panel of figure~\ref{fig:SDSS_HSCz4_zsp_gr}, we plot the color and
magnitude distributions of the 534 quasars at $z=3.0-4.5$ from 
the spectroscopic database of the twelfth data release (DR12) of the SDSS
\citep{Alam2015}
in the S16A-Wide2 coverage. They are within the coverage and neither masked
nor flagged by the masking process described in sections~\ref{sec:database} 
and \ref{sec:mask}, except for the HSC $i<22$ mask.
Red open and blue filled circles
represent quasars which are recovered and missed in the above selection, respectively, 
with the photometric data of the HSC.
Because the bright end of the sample can be affected by 
saturation or non-linearity, figure~\ref{fig:SDSS_HSCz4_zsp_gr} is 
plotted with the colors in the SDSS photometry
converted to the HSC system with the conversion discussed in section 3.3.
Quasars with red $g-r$ color should be selected with the color
selection, but they are missed above around $i<19.8$ mag. Although
in this plot, we remove objects which have saturated pixel at
the central $3\times3$ pixels by the flagging discussed in 
section~\ref{sec:database}, still some bright $i<20.0$ mag stellar objects seem 
to be affected by saturation or non-linearity effects. 
Therefore, hereafter we limit the 
sample fainter than $>20.0$ mag in all of the $r$, $i$, $z$, and $y$ bands, 
for statistical discussions.
There are 379 objects fainter than $i>20.0$ mag 
among the 534 objects.

In the right panel of the figure, we plot $g-r$ color as a function 
of spectroscopic redshift of the 379 SDSS quasars. 
The $g-r$ color of the quasars at $z>3.5$ are red, and most of them 
are selected by the above color selection criteria;
92 quasars are above $z=3.5$ and 61 are selected by the above 
$z=4$ quasar selections, i.e. 66\% of the SDSS quasars are selected.
Among the 31 missed quasars, 1, 26, and 4 objects are missed by 
the first stellarity, the second $g-r$ vs. $r-z$ color-color, and
the third $i-z$ vs. $z-y$ color-color selections, respectively.

After applying the stellarity and color criteria to the
$20.0<i<24.0$ objects which meet the flag conditions in table~\ref{tab:TIDYselection},
we select 3,227 candidates of $z\sim4$ quasars.
The number of candidates in each sub-region is summarized in 
the column $N_{\rm org}$ of table~\ref{tab:subregions}. 
However, if we check the 5-band
images of the selected candidates, we find that junk objects contaminate
the sample. Therefore, we apply further masking based on 
the mask image and the bright star catalogs as described below.

% \begin{figure}
%  \begin{center}
%   \includegraphics[scale=0.5]{ZSPstellar_HSCWIDE_zsphist.eps}
%  \end{center}
% \caption{
% Redshift histogram of HSC stellar objects with 
% spectroscopic information. Objects meeting the 
% $z\sim4$ quasar selection criteria are shown with red histogram.
% \label{fig:ZSPstellar_HSCWIDE_zsphist}}
% \end{figure}

\subsection{Additional masking and masking junk objects}
\label{sec:mask}

In addition to the flags listed in the database
as shown in table~\ref{tab:TIDYselection},
we further check similar flags in the mask images
provided in the stacked image database.
The primary aim of this further masking process
is to remove the junk objects. Additionally, we
try to easily mimic the flag conditions shown in the 
table~\ref{tab:TIDYselection} when we construct
mock random objects uniformly distributed 
in the survey region, which is necessary in the
evaluation of the effective survey area 
(see section~\ref{sec:survey_area}).
We mask objects which have either of the 
MP\_EDGE, MP\_BAD, MP\_SAT,
MP\_NODATA, or MP\_NOT\_DEBLENDED
flag within the radius of \timeform{5"} and
MP\_CR in the central $3\times3$ pixels.
These masking is usually stricter than
the flag in table~\ref{tab:TIDYselection}, 
therefore by considering these masking in the
process of the random mock objects, we expect
all of the flagging and masking conditions are
simulated.

The above flags and masking are not sufficient to remove
objects with bad photometry in the catalog. For example, 
satellite tracks remaining in the stacked images
can be cataloged as objects detected in one band.
Such tracks themselves are not selected as a $z\sim4$ quasar
candidate, but affect the photometry of real objects 
close to them.
Ghost images and faint halo around bright stars 
can also be cataloged as objects meeting the
above flag conditions, and also affect the photometry
around them.

At first, 
we remove objects around bright stars and galaxies.
In the HSC catalog, objects around bright stars
are flagged based on 
Naval Observatory Merged Astrometric Dataset \citep{Zacharias2005}. 
We remove objects
that have MP\_BRIGHT\_OBJECT flag within the radius
of \timeform{5"} in the mask image. The process 
should mimic flags\_pixel\_bright\_object\_any flag 
in the database.
In addition to the flag, we further consider bright 
objects cataloged in Guide Star Catalog (GSC) version 2.3.2.
We remove objects within $r_{\rm mask}[^{\prime\prime}]=150.0$ from
stars brighter than 10 mag, and $r_{\rm mask}$[$^{\prime\prime}$]
$=20.0+17.1\times(13.5-{\rm mag})$
from stars down to 15 mag. Among the multiple magnitudes 
available for one object in the GSC, we refer the brightest 
magnitude of the object in the above calculation.
The size of the 
additional masks are empirically
determined by checking distribution of "junk" objects
around bright stars.

We also remove objects close to satellite
tracks and faint halos around bright stars, because
photometries of such objects are often unreliable.
In the current database, such
satellite tracks and faint halos
are detected as a widely extended object, and
the pixels associated with them are flagged
as MP\_DETECTED in the mask image same as for
other real astronomical objects. We check $10^{\prime\prime}\times10^{\prime\prime}$
region around each cataloged object and if more than 60\% of the pixels
around a object are flagged as MP\_DETECTED in either of the 5 bands, 
the object is removed. Additionally, we examine the connected
pixels around the object with this flag and if the total number of the 
connected pixels is more than 30\% of the $10^{\prime\prime}\times10^{\prime\prime}$ pixels, 
we also mask the object. 
Furthermore, if the number of detected pixels in the 
$i$ band is 2.0$\times$ larger than that in either of the $r$ or $z$ band, 
or those in the $g$, $r$, $z$, $y$ bands are 2.0$\times$ larger than the $i$ band,
we mask the object.
The size and fraction of the mask
are determined by checking objects around remaining satellite tracks and
faint halos.

The deblending process of the HSC pipeline
does not provide reliable photometry for faint objects in 
the outskirts of relatively bright galaxies ($i<21$).
The above masks around very bright objects are not
sufficient to remove such faint objects around
bright galaxies. Therefore, we remove objects around
stars and galaxies brighter than $i<22.0$ mag. The radius of the
mask, $r_{\rm mask}$, is determined with the adaptive moment of the object 
with $r_{\rm mask}=1.8\times \sqrt{\rm i\_hsm\_moment\_11}$.
In order
to recover real candidates with $i<22.0$ mag, we only
consider the mask for objects fainter than $i>22.0$ mag.

In the selection process, we apply all of the above
additional masking processes after selecting 3,227 candidates of $z\sim4$
quasars with the stellarity and color selections described
above. Once we apply the above masking processes, there are
1,668 candidates left. The number in each sub-region is summarized
in table~\ref{tab:subregions}.
We check the selected candidates by eye, and confirm
almost all of the junk objects and objects in the outskirt
of bright objects are removed. We do not apply a junk object
removal with the eye-ball check. The same masking processes
are also applied to random mock objects described in 
section~\ref{sec:survey_area} to mimic the object detection process.

\section{$z\sim4$ quasar number counts}

\subsection{Modeling photometric uncertainty at each position}

In order to derive the number counts of the 
$z\sim4$ quasar sample, we evaluate the effective survey
area of the S16A-Wide2 dataset as a function of magnitude
and redshift. The effective survey area of the
S16A-Wide2 dataset
needs to be evaluated with considering the variation
of the depth of the images due to the variation
of seeing and atmospheric transparency conditions 
and the number of available
images at each location. At first, we establish
a method to evaluate uncertainty at each position based on 
the variance images of the stacked images.
Then, we construct libraries of quasar photometric models, 
randomly locate the quasar models within the survey region, and
add random photometric error evaluated with the
variance value at the random position.
In this process, we assume that the photometric errors
are dominated by the background noise. Finally, 
we apply the same magnitude and color selection 
to the random quasar models and evaluate effective
survey area by the ratio of recovered $z\sim4$ quasar models
to total input models at each redshift and magnitude.

The photometric uncertainties
of real objects correlate well with the variance
values of the object positions. The correlation
shows the dependence on the size of the object. Therefore, we
construct equations to calculate photometric
uncertainties based on the model PSF size and the
variance at each position. In this uncertainty
model, it is assumed that photometric uncertainty
is not dominated by the object Poisson noise. In reality
the objects at the detection limits are faint enough
to meet the assumption.

\subsection{Constructing a library of quasar photometric models with templates}
\label{sec:SEDlib}

The quasar photometric models are constructed based on 
a library of spectral energy distributions
(SEDs) of $z\sim4$ quasars, which describes the distribution 
of the SEDs of population of broad-line quasars.
We make a library of quasar SEDs considering the scatter of the
power-law slope, the broad-line equivalent width (EW), 
and strength of absorption by the inter-galactic medium (IGM).
We assume quasar SEDs depend on luminosity, but do
not depend on redshift.

The library of the quasar SEDs is constructed in the
same way as described in \citet{Niida2016}. The underlying
power-law component, $f_{\nu}\propto\nu^{\alpha}$ is 
modeled with three components covering different wavelength
ranges. Below 1100 {\AA}, we use $\alpha$ of $-1.76$ 
following \citet{Telfer2002}. We apply $\alpha$ of
$-0.46$ and $-1.58$ between 1100 {\AA} and 5011 {\AA} and
above 5011 {\AA}, respectively, following
\citet{VandenBerk2001}. For the two power-law
components below 5011 {\AA}, we assume a scatter around the 
average $\alpha$ with standard deviation of 0.30
following \citet{Francis1996}.

The strength of emission lines are modeled in relative
to the C\emissiontype{IV} emission line following the
emission line ratios tabulated in \citet{VandenBerk2001}.
The strength of the C\emissiontype{IV} emission line
is modeled by considering the dependence of the
EW on quasar luminosity, the so-called Baldwin effect.
We do not consider the luminosity dependence of the
emission line ratios.
The average and standard deviation of the C\emissiontype{IV} emission 
line EW are measured as a function of luminosity from 
quasar spectra of the Baryon Oscillation Spectroscopic Survey (BOSS) 
in the SDSS III. The measurement covers
the luminosity range of $M_{\rm 1450}=-21.5$ mag to $-29.5$ mag with 
1 magnitude bin \citep{Niida2016}. In addition to the 
isolated emission lines, the Fe\emissiontype{II} multiplet
emission features and Balmer continuum are added
by using the template in \citet{Kawara1996}. 

Finally we apply the effect of the absorption by
the IGM. We use the updated number density of Ly$\alpha$
absorption system in \citet{Inoue2014}. We also consider
the scatter of the number density by a Monte Carlo
method described in \citet{Inoue2008}.

\begin{figure*}
 \begin{center}
  \includegraphics[scale=0.60]{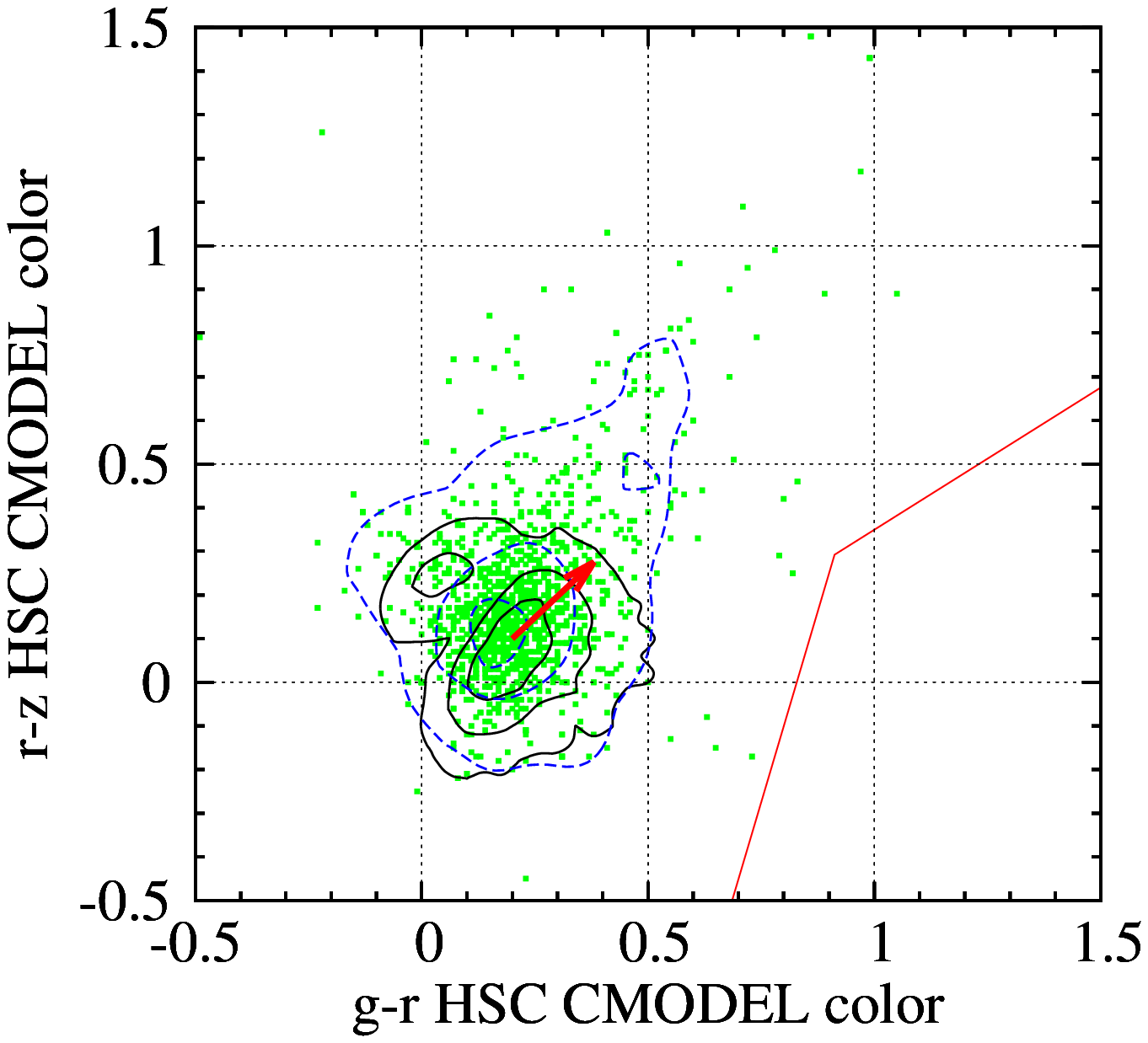}
  \includegraphics[scale=0.60]{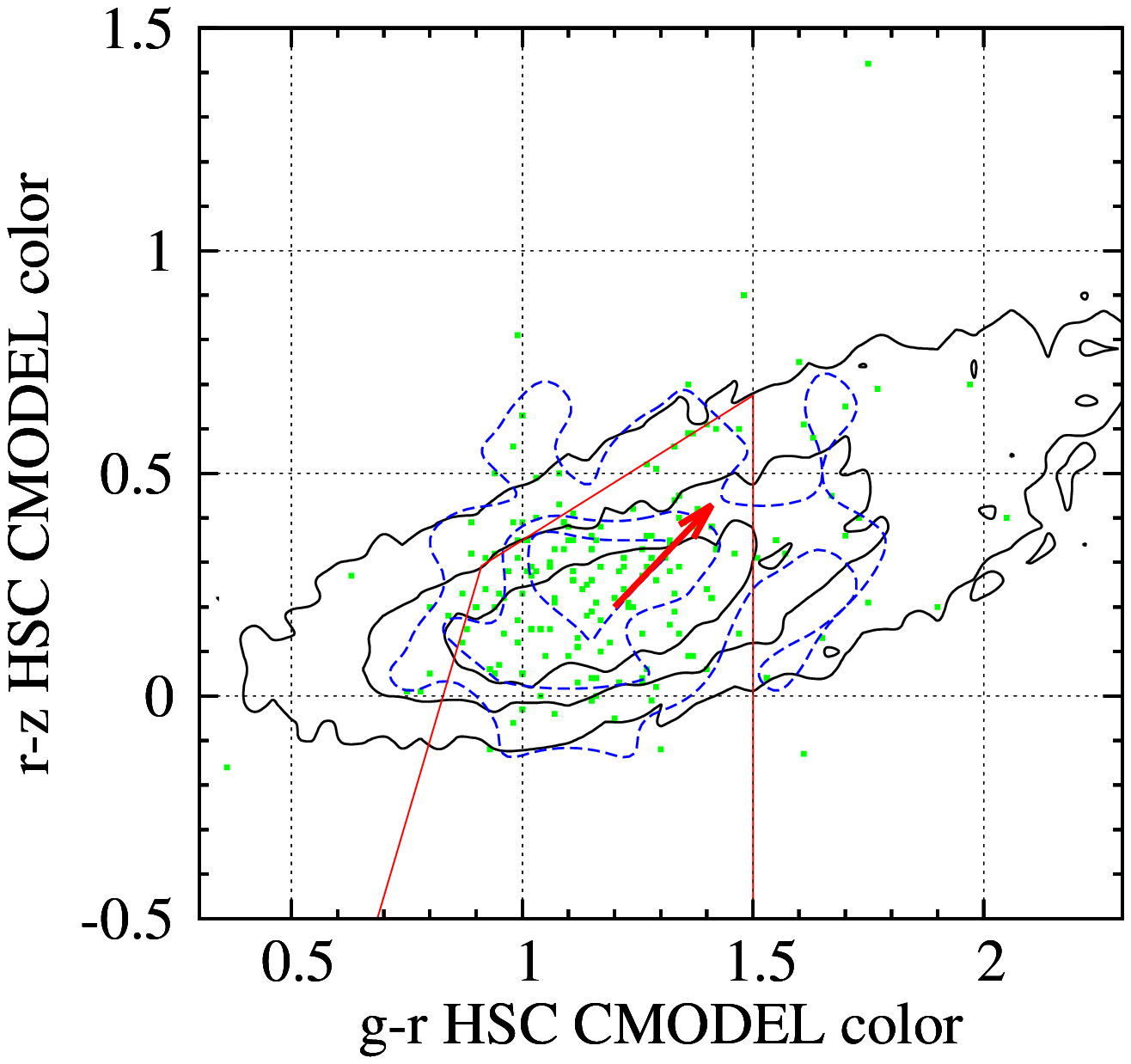}
 \end{center}
 \caption{
Distribution of the model quasars in the $g-r$ vs. $r-z$ color-color
plane (solid contour) compared to SDSS quasars with $i=20-22$ mag within the
S16A-Wide2 coverage. Green point indicate colors of each SDSS quasar, and
dashed contour show the distribution.
Left and right panels
are for quasars at $z=2.5-3.0$ and $z=3.5-4.3$, respectively.
There are 1407 and 234 SDSS DR12 quasars in the redshift and magnitude
ranges in the S16A-Wide2 coverage.
In order not to be affected by the difference in the absolute
magnitude, only model quasars in the same magnitude range are plotted.
The contours represent 30\%, 60\%, and 90\% enclosing area.
Red arrow indicate the effect of SMC-like dust-extinction with
$E(B-V)=0.04$ mag.
Red solid lines represent the $z\sim4$ quasar selection criteria
on the plane. It should be noted that the photometric uncertainty of
the HSC is negligible for objects brighter than the SDSS spectroscopy.
\label{fig:QSOtemplate_SDSSQSO_gr_rz}}
\end{figure*}

We construct 1,000 quasar SEDs in each magnitude bin
from $M_{\rm 1450}=-21.5$ mag to $-29.5$ mag. Therefore
in total 8,000 quasar SEDs are constructed. Each spectrum
is redshifted, and converted to observed flux. 
In figure~\ref{fig:QSOtemplate_SDSSQSO_gr_rz},
the distributions of the model quasars in the redshift ranges
$z=2.5-3.0$ and $z=3.5-4.3$ are plotted with contours
in the left and right panels, respectively. The contours 
enclose 30, 60, 90\% of the models in the redshift range
from top to bottom. We only consider model quasars with apparent
magnitude between $i=20.0$ to $22.0$ mag. The distributions
are compared to those of the SDSS DR12 quasars which are 
in the same magnitude range covered by the HSC S16A-Wide2.

The 60\% enclosing contour of the SED library reproduces
that of the SDSS $z=2.5-3.0$ quasars. However, the 90\%
enclosing contour of the SDSS $z=2.5-3.0$ quasars shows
extended distribution toward red $r-z$ color.
Dust reddening to the quasar spectrum can explain the
extended distribution. The direction of the extension is
consistent with the reddening vector with Small-Magellanic
Cloud-like dust extinction; the red arrow in the figure show
the reddening vector with $E(B-V)=0.04$ mag.
\citet{Richards2003} report 6\% of SDSS
quasars show red color, which is explained with 
$E(B-V)$ larger than 0.04 mag. After correcting for the
bias against reddened quasars, they estimate total 
contribution of such dust-reddened broad-line quasars in the total 
quasar population as 15\%. 

In the current SED library, 
we do not include such population of quasars with 
dust reddening, primarily because current quasar samples
at higher redshifts do not cover such a population
(e.g. \cite{Richards2006}, \cite{Ross2013}; \cite{McGreer2013}).
In the higher redshift range, $z=3.5-4.3$, 
the $r-z$ color distribution of the SDSS quasars is well
reproduced by the distribution of the quasar SED library. 
That means current sample of $z=3.5-4.3$ SDSS quasars does
not include the reddened quasar population seen at $z<3$
or the fraction of such reddened quasar is really decreasing. 
In the current photometric sample, it is hard to tackle 
such reddened quasar population due to heavy contamination
by Galactic stars, and we need future wide-field multi-wavelength
surveys to pick up the population.

On the other hand, $g-r$ color distribution of the SDSS quasars
is narrower than that of the model quasars. The $g-r$
color distribution in this redshift range is mostly determined
by the strength and scatter of the IGM absorption.
Currently, the number of $z=3.5-4.3$ SDSS quasars covered
by HSC photometry is rather limited to disclose the range of the
color distribution, therefore we use the color distribution 
of the model quasars as a baseline sample of the full quasar
population in this paper. Hereafter, we refer to the library of the 
quasar models as the library of quasar photometric models
with templates.

It should be noted that the quasar SED library is 
constructed over a wide luminosity range using the spectra of
less-luminous quasars at lower redshifts, assuming 
that the spectral shapes of the quasars do depend on 
luminosity, but do not depend on redshift.
Therefore, the library extends to fainter luminosity 
range than the SDSS sample at $z=4$ and is therefore applicable
in this HSC study. 

\subsection{Quasar photometric models with a SDSS photometric sample}

The other library of quasar photometric models is constructed by converting
quasar PSF photometry from the SDSS DR12 quasar catalog into the 
HSC photometric
system. Because the HSC survey area is not wide enough to
cover large numbers of SDSS quasars at high-redshifts, and 
quasars brighter than $i<20.0$ mag are affected by
saturation in the HSC photometry, we construct a library
of model quasars by converting SDSS photometry to HSC photometry
instead of directly using the HSC photometry of the
SDSS quasars.

We derive the conversion by applying filter 
response curves of the HSC \citep{Kawanomoto2017} and SDSS
\footnote{\texttt{http://classic.sdss.org/dr7/instruments/imager}} system to 
the spectrophotometric library of 
Galactic stars \citep{Gunn1983}.
The conversion in each band is determined 
by a linear dependence on one color, and they are

\begin{eqnarray}
 g_{\rm HSC} &=& g_{\rm SDSS} - 0.074 (g_{\rm SDSS} - r_{\rm SDSS}) - 0.011 \\
 r_{\rm HSC} &=& r_{\rm SDSS} - 0.004 (r_{\rm SDSS} - i_{\rm SDSS}) - 0.001 \\
 i_{\rm HSC} &=& i_{\rm SDSS} - 0.106 (r_{\rm SDSS} - i_{\rm SDSS}) + 0.003 \\
 z_{\rm HSC} &=& z_{\rm SDSS} + 0.006 (i_{\rm SDSS} - z_{\rm SDSS}) - 0.006 \\
 y_{\rm HSC} &=& z_{\rm SDSS} - 0.419 (i_{\rm SDSS} - z_{\rm SDSS}) + 0.030. 
\end{eqnarray}

Although these equations are determined for the
SEDs of Galactic stars, they are effective to
quasars. If we compare the HSC photometry of SDSS quasars
with the SDSS photometry converted with the above
equations, converted SDSS colors of SDSS quasars are 
consistent with their colors measured in the HSC dataset
within a scatter of $\sim0.2$ mag without
any systematic offset. It needs to be noted that
the above equations are applicable only to 
smooth SEDs and may not work with SEDs with a
break or strong emission line.

\begin{figure*}
 \begin{center}
  \includegraphics[scale=0.60]{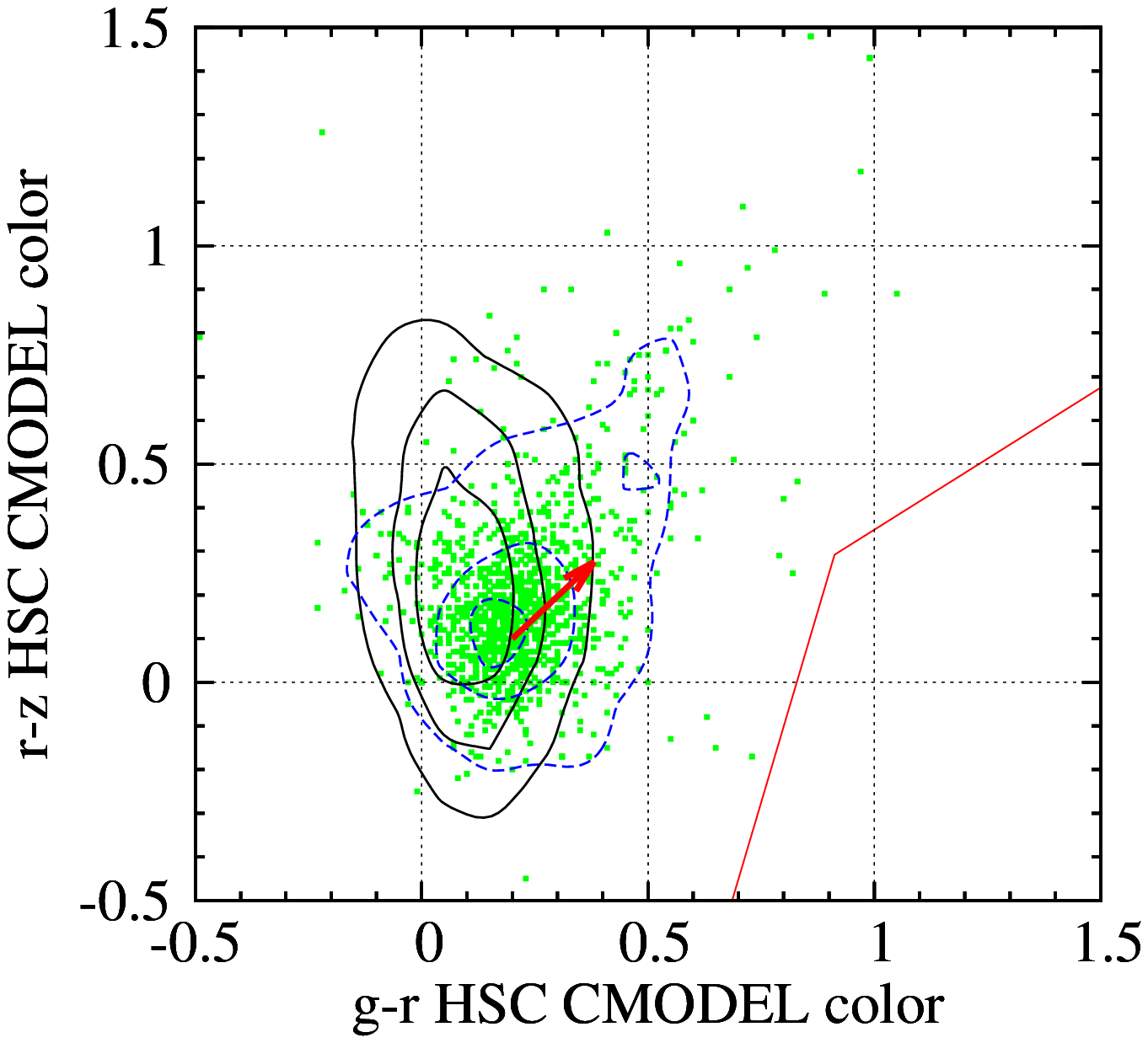}
  \includegraphics[scale=0.60]{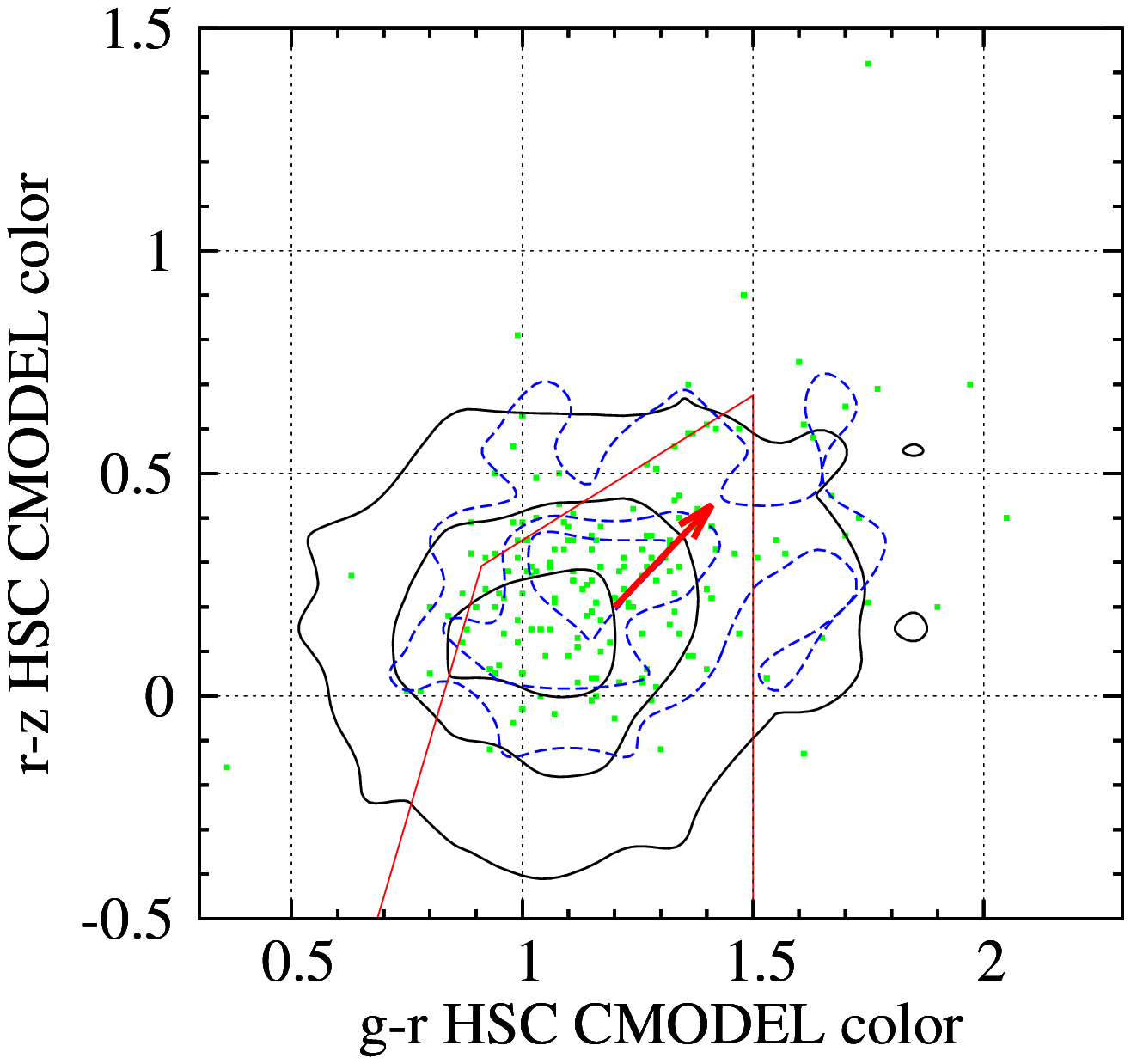}
 \end{center}
 \caption{
Distribution of the quasar photometric models with SDSS photometry
in the $g-r$ vs. $r-z$ color-color plane
is shown with the solid contour. SDSS photometry is converted to the HSC system.
Green point indicate HSC
colors of SDSS quasars with $i=20-22$ mag within the
S16A-Wide2 coverage with dashed contour showing their distribution.
Left and right panels
are for quasars at $z=2.5-3.0$ and $z=3.5-4.3$, respectively.
There are 43123 and 4522 quasars in the redshift and magnitude
ranges in the SDSS DR12 quasar catalog.
The contours represent 30\%, 60\%, and 90\% enclosing area.
Red arrow indicate the effect of SMC-like dust-extinction with
$E(B-V)=0.04$ mag.
Red solid lines represent the $z\sim4$ quasar selection criteria
on the plane.
\label{fig:SDSSmock_SDSSQSO_gr_rz}}
\end{figure*}

In figure~\ref{fig:SDSSmock_SDSSQSO_gr_rz}, we compare the
distribution of the model quasars with SDSS photometry 
at $z=2.5-3.0$ and $z=3.5-4.3$ with that of the SDSS quasars
observed by the HSC system in the magnitude range $i=20-22$ mag.
The distribution of the converted photometry
is broadly consistent with the HSC colors of SDSS quasars directly 
measured in the HSC images. For quasars at $z=2.5-3.0$, the $g-r$
colors of the quasars with red $r-z$ colors show systematically
bluer distribution. Hereafter, we refer to the library as
the library of quasar photometric models with SDSS photometry.

\subsection{Estimating survey area as a function of redshifts and magnitudes}
\label{sec:survey_area}

The effective survey area is determined as a function of 
magnitude and redshift for the quasar sample.
We use the above 2 libraries of quasar models to
derive the effective survey area correcting the fraction of missed
quasars among the "complete" sample. In order to derive the survey
area at a certain magnitude limit, firstly random positions are
selected within the survey region, then we apply the same masking
processes described in section~\ref{sec:mask}. We pick-up 6,000
random positions per deg$^{2}$. The concept of the random 
positions is the same as what is available in the database as
the random objects \citep{Aihara2017b}. Secondly we randomly
assign one quasar model from one of the libraries.
For the quasar models with template, we randomly draw a
quasar model in the considered magnitude range,
and for the quasar models with SDSS photometry, 
we randomly select one quasar, and normalize the
SED to match the considered magnitude.
Therefore, in the calculation with the quasar models with
SDSS photometry, we ignore
the luminosity dependence of the quasar SEDs.
Then, we calculate uncertainties associated with the photometry in 
each band and we apply random fluctuations to the photometric model.
Finally, we apply the magnitude ($20.0<i<24.0$) and color
selection criteria to examine the fraction of recovered objects
at the magnitude.

\begin{figure*}
 \begin{center}
  \includegraphics[scale=0.95]{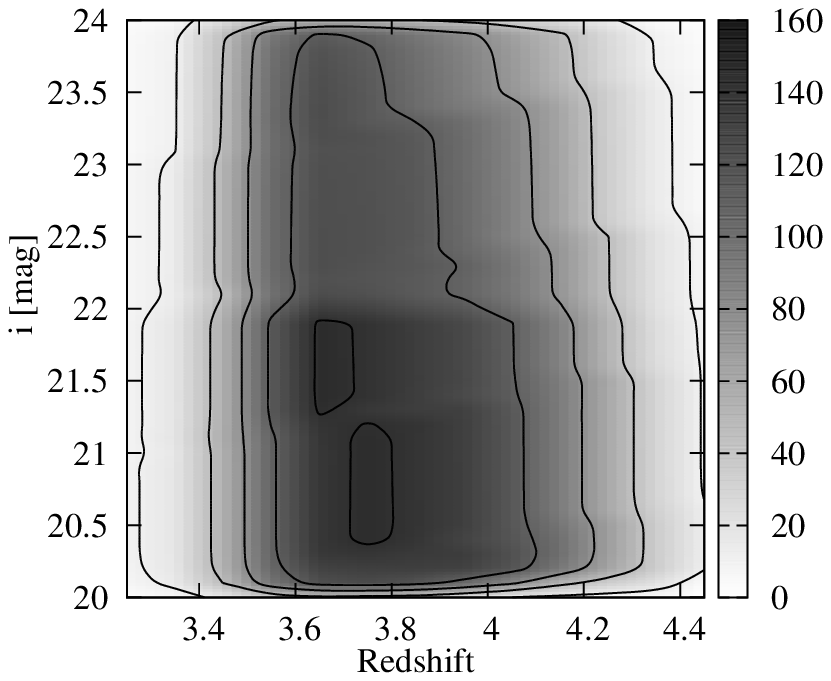}
  \includegraphics[scale=0.95]{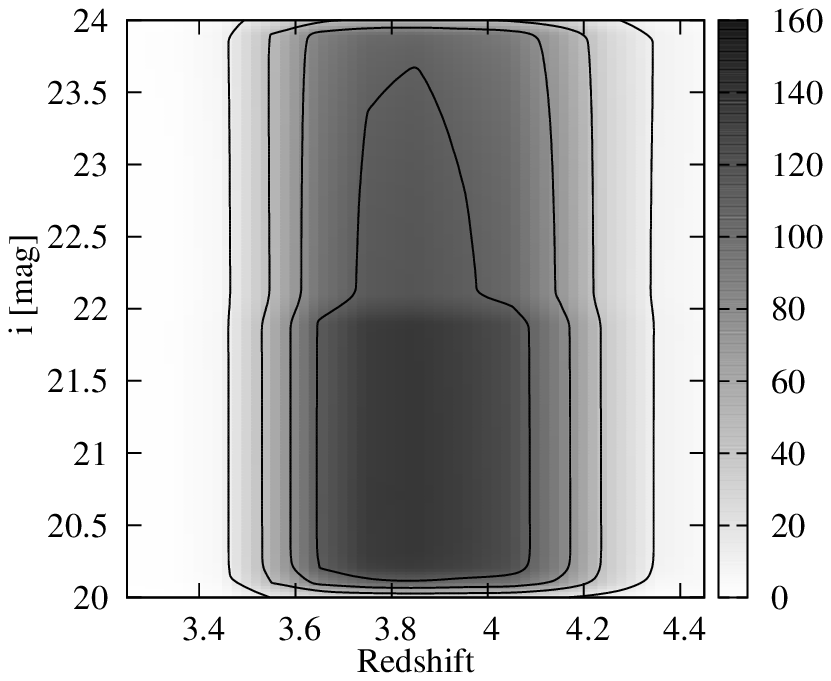}
 \end{center}
\caption{
Survey area as a function of redshift and $i$-band magnitude.
Left) derived with the quasar templates. 
Right) derived with the SDSS quasar photometry.
The contours are plotted at 
10deg$^{2}$, 40deg$^{2}$, 70deg$^{2}$, 100deg$^{2}$, and 130deg$^{2}$.
\label{fig:WIDE_area_z4QSO}}
\end{figure*}

The resulting effective survey areas as a function of redshift at
each magnitude are shown in figure~\ref{fig:WIDE_area_z4QSO}.
The left and right panels show the effective areas estimated 
with the quasar models with templates and those with SDSS photometry, respectively.
They are broadly consistent with each other, though the area
derived with the models with templates shows a broader redshift distribution
than that with the models with SDSS photometry as expected from the color
distributions shown in figure~\ref{fig:QSOtemplate_SDSSQSO_gr_rz}.
Both areas show a break at $i=22.0$ mag, caused by the
mask around objects with $i<22.0$ mag. The peak of the survey
area is 133.0 deg$^{2}$ for $i=21.4$ mag quasars at $z=3.65$.
In table~\ref{tab:subregions}, effective areas after removing
masked regions are summarized. The effective areas are calculated
applying the same masking process described in section
\ref{sec:mask}. For the 6-th and 7-th columns, the
areas without and with the masks with objects brighter than $i<22$ mag, 
respectively, are shown. The total effective survey area for $i<22$ mag
objects is 172.0 deg$^{2}$. It should be noted that the effective
area for $z\sim4$ quasars 
includes not only the effect of the
masked regions and shallow areas, but also the selection efficiency
of the quasars at each redshift.
The peak of the survey area for $z\sim4$ quasar 
is 77\% of the total effective survey
area for $i<22$ mag objects. The fraction is broadly consistent
with the fraction of $z>3.5$ SDSS quasars meeting the $z\sim4$ quasar
selection criteria (66\%) as discussed in section~\ref{sec:colorsel}.

\begin{figure}
 \begin{center}
  \includegraphics[origin=c,angle=-90,scale=0.80]{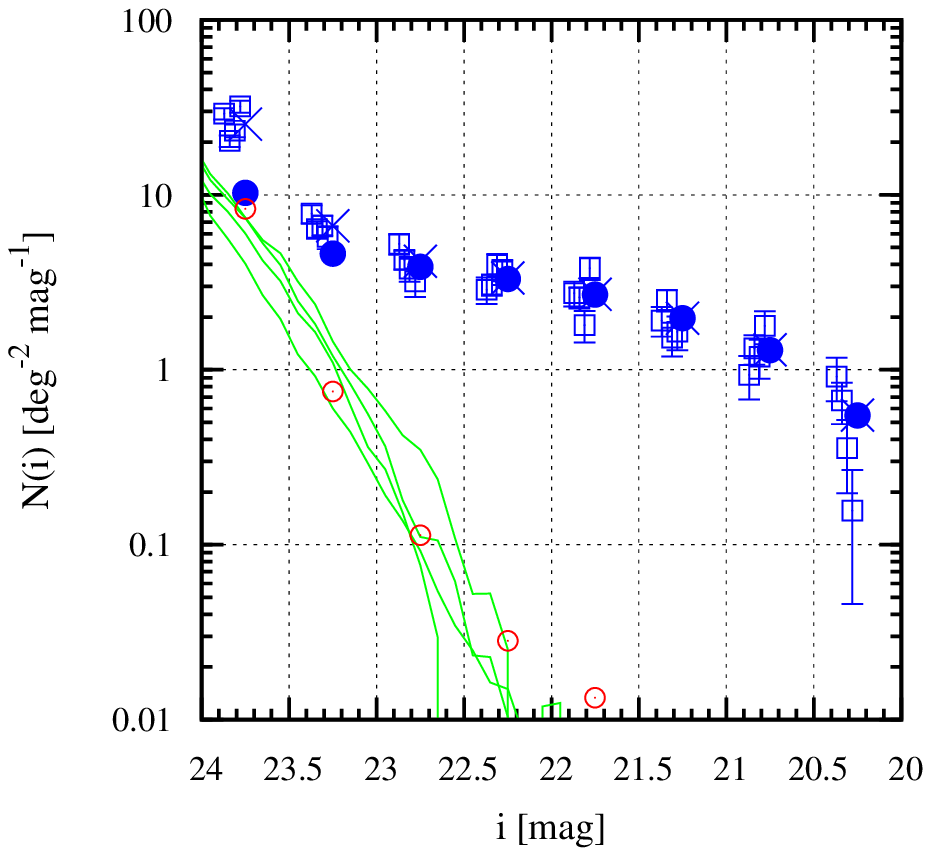}
 \end{center}
\caption{
Differential number counts of $z\sim4$ quasar candidates.
Blue crosses represent the number counts derived 
in the entire S16A-Wide2 region. 
Open squares are the number counts 
after dividing the region into the 4 sub-regions. 
The uncertainties associated with the number count in 
each sub-region are
determined with the Poisson statistics. 
The filled circles show the number counts after correcting
for the expected number of contamination by Galactic stars
(green solid line for each sub-region) and the compact galaxies 
(red open circles). The open squares in each sub-regions are
shifted in horizontal direction for clarity.
\label{fig:WIDE_z4QSO_num}}
\end{figure}

The differential number counts of the $z\sim4$ quasars are shown
in figure~\ref{fig:WIDE_z4QSO_num} and summarized in table~\ref{tab:z4number}. 
The second and third columns show the raw number and surface number
density of the quasars, respectively.
We apply the average effective survey area between $z=3.6-3.8$, where
the color selection efficiency is maximized. In the number counts, 
we correct for the incompleteness of the stellarity selection
assuming the median condition in figure~\ref{fig:HSCUDEEP0_STAR11_ACSstellar_frac}.
We do not correct for the contamination by the extended objects, 
because we consider the contamination in the next sub-section.
The number counts show steady
increase toward the faint-end, and excess at magnitudes fainter than $i>23.0$ mag.
We also examine the number counts in each sub-regions. The results are
shown with open squares with error bars estimated with the Poisson 
statistics. The number counts are consistent with those derived
in the entire region, and the effect of the cosmic variance seems to be small,
thanks to the large survey area. The scatter at the brightest end is large
due to the limited number of the sample.

\subsection{Contamination by extended objects}
\label{sec:cont_extend}

The expected number of contamination by extended objects
that meet the $z\sim4$ quasar color selection criteria is
evaluated. In addition to the Lyman Break Galaxies at $z\sim4$,
foreground $z<1$ galaxies can be contaminants to the
$z\sim4$ quasar sample, because they can have similar
colors to $z\sim4$ quasars due to their 4000 {\AA} break.
It should be noted that even with the ground-based HSC
images, most of the $z\sim4$ LBGs are extended,
thanks to the good image quality in the $i$ band,
thus only compact LBGs are contaminating the stellar object
selection.

\begin{figure}
 \begin{center}
  \includegraphics[origin=c,scale=0.60]{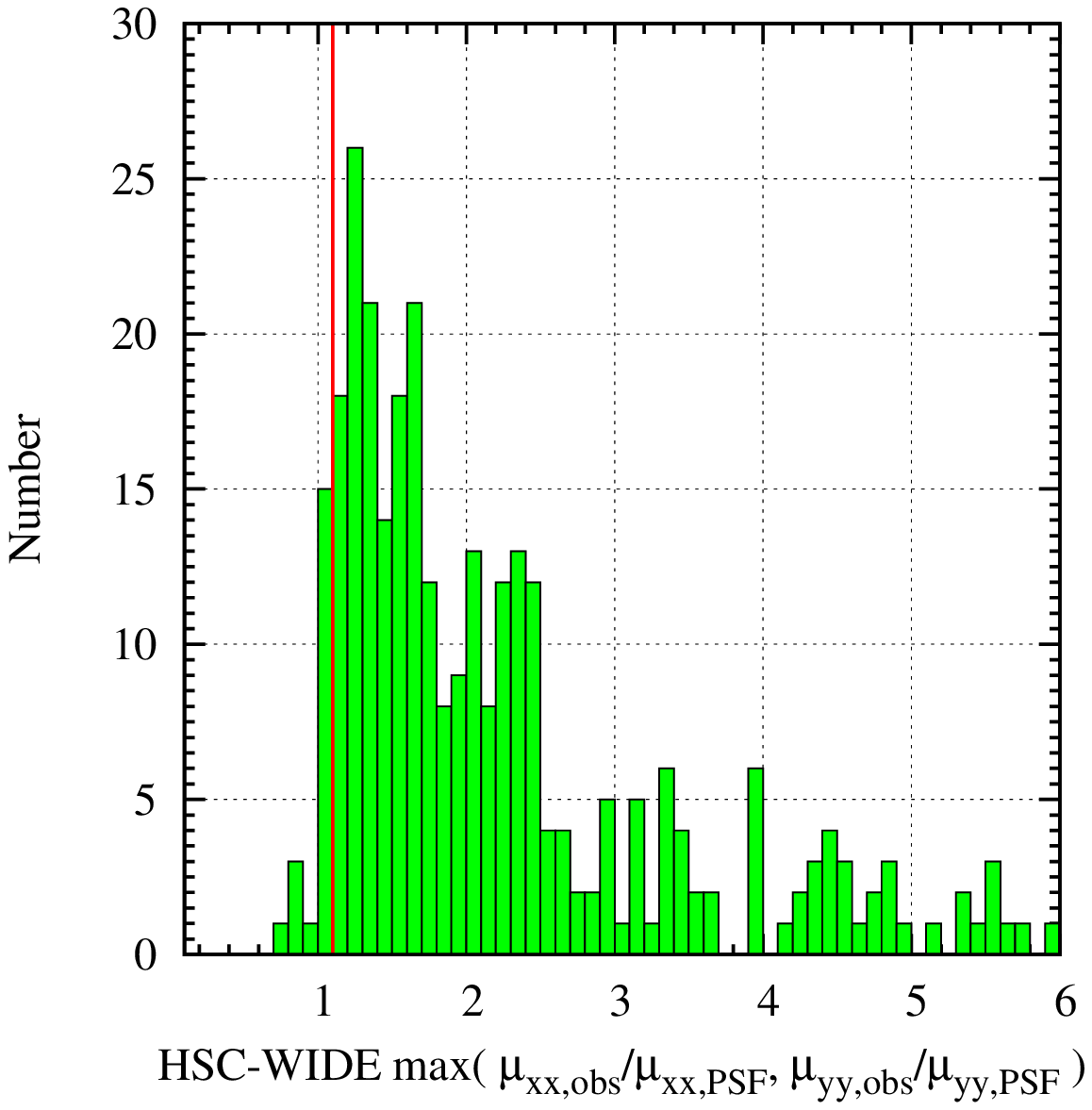}
 \end{center}
\caption{
Distribution of the measured adaptive moment ratios
of the 306 $z>3$ $i>23$ galaxies with spectroscopic identification.
The red solid line indicate the selection criteria for
stellar objects.
\label{fig:HSCWIDE_z4spec_HSMshape_hist}}
\end{figure}

In figure~\ref{fig:HSCWIDE_z4spec_HSMshape_hist}, 
the distribution of the measured adaptive moment ratios 
in the S16A-Wide2 dataset of the 306 $z>3$ and $i>23$ galaxies that
are spectroscopically identified in deep surveys.
Broad-line AGNs are removed in this plot.
The red line indicate the selection criteria of the
stellar objects. Thanks to the good image quality of
the $i$-band image of the Wide layer dataset, they are
significantly extended compared to the stellar objects. 
The fraction of the galaxies that are classified as
stellar is 6\%. The fraction is larger than that 
observed among all galaxies with $i=24$ mag (2\%), 
in this evaluation we use the distribution of all galaxies
in each magnitude range, 
because it is possible that the spectroscopically identified
sample can be biased toward compact galaxies, and we
do not know the true size distribution of $z>3$ galaxies.

At first, we construct a catalog of extended
objects that meet the color criteria from the HSC-SSP S16A-Wide2
database. Then, we apply the masking processes described 
above. The number counts of the object without flag is
calculated with the same survey area used for the
$z\sim4$ quasar number count. The expected number counts of
contamination by extended objects is calculated by 
multiplying the fraction of
ACS extended objects classified as stellar in HSC criteria
as a function of magnitude. The fraction is determined
in the same way as described in section~\ref{sec:SGclass},
but this time the fraction of HSC stellar objects among 
the ACS extended objects is calculated. We use the results
with the median condition.
In this calculation, we assume that the contaminating
extended galaxies have the same size distribution as
the entire galaxies in the same magnitude range.
The resulting number of contamination by extended objects
is shown with
red open circles in figure~\ref{fig:WIDE_z4QSO_num}.
The estimated result suggests that the contamination 
by extended objects can contribute to the number counts
below $i=23.5$ mag, and one third of the $z\sim4$ quasar
candidates in the $i=23.5-24.0$ mag bin can be contamination 
of extended objects, which are non-AGN galaxies.
The estimation is consistent with the rapid increase
of the contamination of extended objects shown in 
figure~\ref{fig:HSCUDEEP0_STAR11_ACSstellar_frac}.

\subsection{Contamination by Galactic stars}
\label{sec:cont_star}

The number of contamination by Galactic stars
is evaluated by multiplying the observed number counts of Galactic stars
in each sub-region with the fraction of Galactic stars
meeting the $z\sim4$ quasar color selection criteria
evaluated as a function of $i$ band magnitude.
Both of the number counts and the fraction are estimated
with a photometric library of Galactic stars in the HSC
system.

\begin{figure}
 \begin{center}
  \includegraphics[scale=0.55]{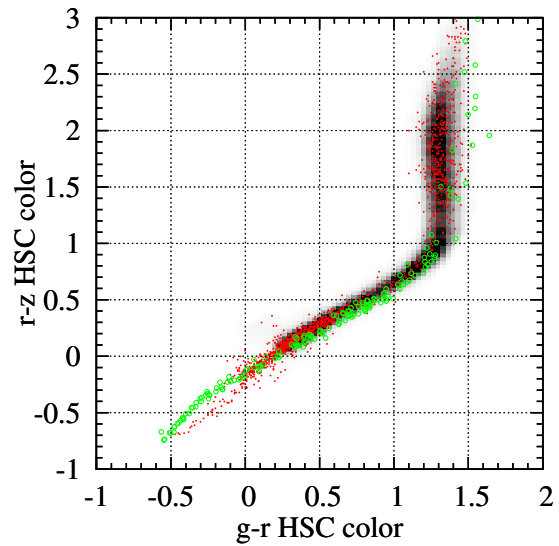}
 \end{center}
\caption{
$g-r$ vs. $r-z$ color-color distribution of Galactic stars.
Red dots indicate the spectroscopically identified Galactic
stars in the S16A-Wide2 database (randomly selected 30\% of stars
are plotted for clarity), and gray scale represent the
distribution of the stellar objects in the S16A-Wide2 database
which are selected based on the distance on the diagram, $r_{\rm col}<0.1$ mag, 
from red dots. Green dots indicate colors of Gunn-Stryker stars 
\citep{Gunn1983} in the HSC system.
\label{fig:HSCSTAR_gr_rz_cont}}
\end{figure}

A photometric library of Galactic stars is constructed as follows.
At first, we collect a list of spectroscopically identified 
Galactic stars in the S16A-Wide2 database which are flagged 
and masked in the same way as described
in sections~\ref{sec:database} and \ref{sec:mask}. Then, we remove some outliers, which are
outside of the stellar sequence seen in the
$g-r$ vs. $r-z$ color-color diagram. The distribution of
resulting stars on the color-color diagram is shown in 
figure~\ref{fig:HSCSTAR_gr_rz_cont} with red dots. This list of stars cover
a wide range of stellar types. However, they can have different
mixture of types from that at the HSC Wide-layer depth
because most of the spectroscopic identifications are from
SDSS DR12 database, whose depth is shallower.
Therefore, we match the list of flagged and masked stellar objects
in the S16A-Wide2 database with the above list of the
spectroscopically identified Galactic stars on the $g-r$ vs. $r-z$
color-color diagram based on the distance on the diagram, $r_{\rm col}$, 
less than 0.1 mag. The distribution of resulting stars are shown 
in the figure with gray scale. It follows the distribution of
the spectroscopically identified stars, but more heavily populated
with late-type stars compared to that of the spectroscopically identified
stars. Open green circles indicate the colors of stars in the
spectro-photometric library of \citet{Gunn1983}. They show systematic
offset from the observed sequence of the late type stars redder than
$r-z>1.0$. Difference in the stellar metallicity is thought to be 
the cause of the systematic offset \citep{Fukugita2011}. 

\begin{figure}
 \begin{center}
  \includegraphics[origin=c,angle=-90,scale=1.10]{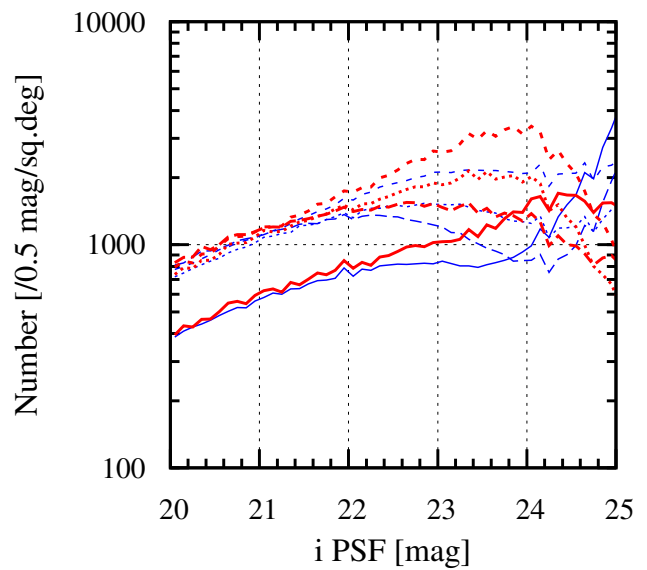}
 \end{center}
\caption{
Differential number counts of Galactic stars in the 4 sub-regions
(solid, dotted, dashed, and long-dashed lines
represent WideA, WideB, WideC, and WideD sub-regions, respectively).
Blue thin lines show raw count, and red thick lines represent after correcting
for contamination by extended galaxies following the contamination
rate shown in figure~\ref{fig:HSCUDEEP0_STAR11_ACSstellar_frac}.
\label{fig:WIDE_STAR_numcount}}
\end{figure}

Then, the number counts of Galactic stars are derived in the 
same way as for the $z\sim4$ quasars. At first, we apply the masking
process described in sections~\ref{sec:database} and \ref{sec:mask} 
to the list of stellar objects in the HSC database. Then
the area of survey region is determined by random model
objects constructed from a photometric library of Galactic stars.
We assume that the photometric error in the photometric library is
negligible, because we only consider stars which have similar colors
to the bright spectroscopically identified stars with 
negligible photometric errors. Then, we add random photometric 
error associated at each location in the same way described 
in section~\ref{sec:survey_area} and pick up random object above 
the detection limit, i.e. $i<24.0$ mag and photometric errors in 
$i$ and $z$ bands are smaller than 0.1 mag.
The resulting number counts are shown in figure~\ref{fig:WIDE_STAR_numcount}.
The raw number counts in each sub-region are shown in thin blue line. 
In order to derive the intrinsic number counts of Galactic stars,
we correct for the incompleteness and contamination of the star-galaxy
separation applying the fraction shown in 
figure~\ref{fig:HSCUDEEP0_STAR11_ACSstellar_frac}. The resulting number count
after the corrections are shown in thick red lines. The number count
shows steady increase toward $i=22-24$ mag, but then shows a plateau
or decline toward the faint-end.

The expected contamination rate to the quasar number counts in 
each sub-region is
shown in figure~\ref{fig:WIDE_z4QSO_num} with the green solid line.
Because photometric uncertainty increases toward
fainter objects, the expected number of contamination of
Galactic stars increases rapidly, although the number counts 
themselves show plateau at $i=22-24$ mag. The expected number
of contamination varies in the 4 sub-regions depending on the distance
from Galactic plane. The contamination by Galactic stars can 
contribute one third of the number counts in the 
$i=23.5-24.0$ and $i=23.0-23.5$ mag bins.

\subsection{Number counts after correcting for the contamination}

\begin{figure}
 \begin{center}
  \includegraphics[origin=c,angle=-90,scale=1.00]{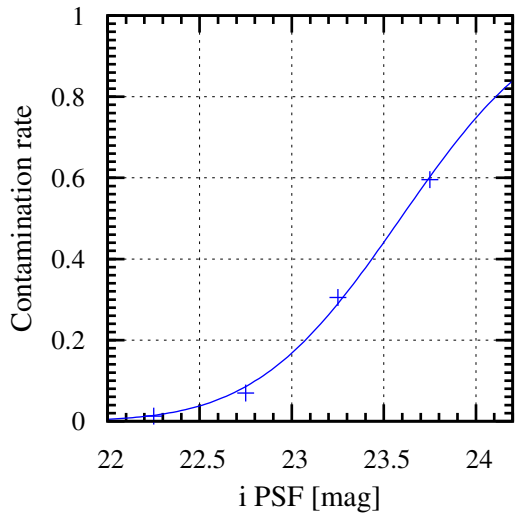}
 \end{center}
\caption{
Contamination rate of the $z=4$ quasar selection. Solid curve
indicates the best fit model with the error function.
\label{fig:WIDE_cont_rate}}
\end{figure}

Based on the estimated number counts from the contamination, we
evaluate the contamination rate, which is the fraction of contaminating
objects among the candidates of $z=4$ quasars.
For the contamination by Galactic stars, the expected number
depends on the Galactic coordinate of a sub-region, we average
the number after weighting the survey area of each sub-region.
The resulting contamination rate in each $i$ band magnitude bin
is shown in figure~\ref{fig:WIDE_cont_rate} with crosses. 
In the magnitude range fainter than $i>23.5$ mag, the contamination
rate is estimated to be higher than 50\%. In order to correct for
the contamination statistically, we fit the contamination rate
with the error function. The best-fit function is derived as
$[p_{\rm cont} = {\rm erfc}(-1.15(i-23.59))/2 ]$, and plotted as
solid line in figure~\ref{fig:WIDE_cont_rate}.

We correct for the expected contamination by assigning weight for
each object based on the probability of non-contamination, i.e. 
the contamination rate subtracted from 1. If the contamination rate
is 0.8 at the magnitude of an object, we assign weight of 0.2 for
the object, and the weight is summed when we calculate the number
count and the luminosity function. The total sum of the weight of
1,666 candidates is 1155.2, i.e. one third of the sample can be
contamination mostly contributing $i>23.5$ mag. The number counts after 
correcting for the contamination are shown in figure~\ref{fig:WIDE_z4QSO_num}
with blue filled circles and summarized in the fourth column of table~\ref{tab:z4number}.
Once we correct for the contamination statistically, the number density shows
monotonic increase toward the faint-end.
The excess seen in the magnitude range $i>23$ mag can be explained
with the contamination by Galactic stars and compact galaxies
that meet the $g$-dropout selection.

\begin{table}
  \caption{Number count of the $z=4$ quasars}\label{tab:z4number}
  \begin{center}
    \begin{tabular}{crcc}
\hline
  $i$         & $N_{\rm obs}$ & $n$                    &  $n_{\rm corr}$   \\
 (mag)        &               & (deg$^{-2}$ mag$^{-1}$) & (deg$^{-2}$ mag$^{-1}$) \\
\hline
$20.00-20.50$ &  34 & $5.50\times10^{-1}$ & $5.48\times10^{-1}$ \\ 
$20.50-21.00$ &  78 & $1.30\times10^{0}$  & $1.30\times10^{0}$ \\ 
$21.00-21.50$ & 121 & $1.97\times10^{0}$  & $1.97\times10^{0}$ \\ 
$21.50-22.00$ & 162 & $2.70\times10^{0}$  & $2.69\times10^{0}$ \\ 
$22.00-22.50$ & 170 & $3.35\times10^{0}$  & $3.31\times10^{0}$ \\ 
$22.50-23.00$ & 189 & $4.17\times10^{0}$  & $3.88\times10^{0}$ \\ 
$23.00-23.50$ & 258 & $6.65\times10^{0}$  & $4.62\times10^{0}$ \\ 
$23.50-24.00$ & 656 & $2.54\times10^{1}$ & $1.03\times10^{1}$ \\ 
\hline
    \end{tabular}
  \end{center}
\end{table}

\section{Results}

\subsection{Redshifts and absolute magnitudes}

\begin{figure}
 \begin{center}
  \includegraphics[scale=0.5]{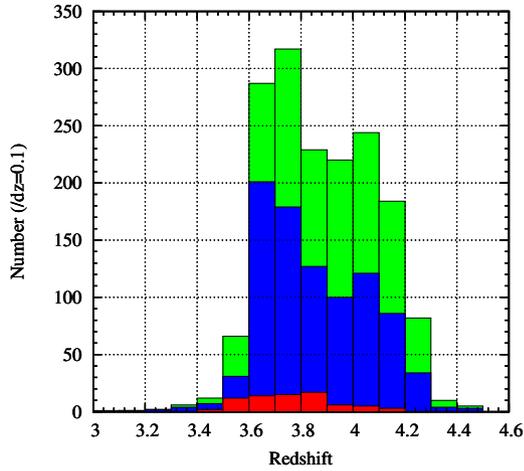}
 \end{center}
\caption{
Redshift distribution of the $z\sim4$ quasar candidates.
Green and blue histograms show the distributions of candidates
with $i<24.0$ and $21.0<i<23.5$, respectively. 
Red histogram indicates the redshift distribution of
quasars with spectroscopic redshifts. 
\label{fig:WIDE_zhist}}
\end{figure}

Among the 1,668 $z=4$ quasar candidates, 76 of them have
spectroscopic redshift information in the literature. 
Most of the redshift identifications come from 
the SDSS quasar surveys. A few of them are 
from deeper surveys (e.g. \cite{Akiyama2015}).
Their redshift
distribution is shown with a red histogram 
in figure~\ref{fig:WIDE_zhist}. They distribute redshift
range between 3.4 and 4.2.
Two SDSS quasars that have redshift $z\sim1$ are contaminating
the sample. The two quasars have $g-r$ colors of 0.37 and 0.50
in the SDSS database, while those in the HSC are 0.81 and 0.95.
Because they are located within \timeform{34'} each other, 
their HSC photometry could be affected by an unknown 
photometric problem in the specific region.

\begin{figure}
 \begin{center}
  \includegraphics[scale=0.85, origin=c, angle=-90]{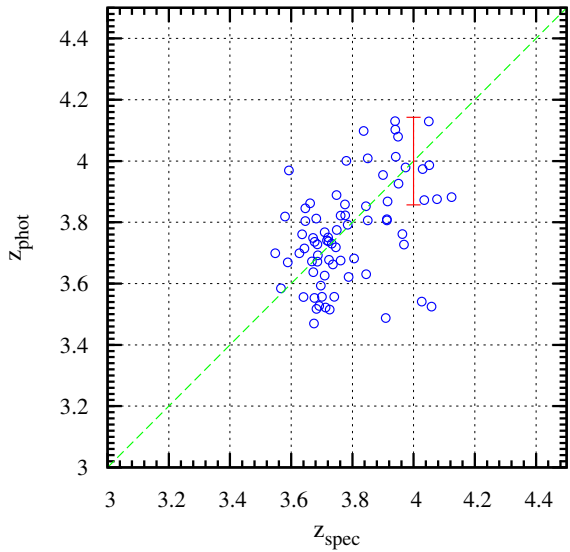}
 \end{center}
\caption{
Photometric redshift vs. spectroscopic redshift. Green dashed line
indicates the equality. The error bar indicates the $\sigma$ of 
the difference between $z_{\rm spec}$ and $z_{\rm phot}$.
\label{fig:HSCz4QSO_zspec_zphot}}
\end{figure}

For the remaining candidates, no spectroscopic
redshift information is available. We derive their
photometric redshifts using the library of 
quasar models with templates
described in section~\ref{sec:SEDlib}. For each
candidate, we calculate $\chi^2$ with all of the
templates in the redshift range between $2.5<z<6.0$
with step of $dz=0.1$.
As a Bayesian photometric redshift estimation,
we take the average of redshift with weighting by
$\exp(-\chi^2/2)$. We assume uniform prior in the 
above redshift range. The resulting photometric
redshift is plotted against the spectroscopic redshift
in figure~\ref{fig:HSCz4QSO_zspec_zphot}.
The derived photometric redshifts correlate well 
with the spectroscopic redshifts.
If we remove 2 outliers whose redshifts are $z\sim1$, 
the average and $\sigma$ of the difference between
$z_{\rm phot}$ and $z_{\rm spec}$ are
are $-0.007$ and $0.143$, respectively. 
The systematic offset between the 
$z_{\rm phot}$ and $z_{\rm spec}$ is negligible.

The distribution of the $z_{\rm phot}$ and $z_{\rm spec}$
is shown in figure~\ref{fig:WIDE_zhist} with green histogram. 
Most of the selected quasars
distribute between $z=3.5$ and $4.3$. The range is 
consistent with the selection function evaluated
in figure~\ref{fig:WIDE_area_z4QSO}. The average of the
redshift of the sample is 3.9. The redshift distribution of
quasars in $21.0<i<23.5$, which are less affected by 
contamination, shows a similar distribution as
the entire sample.

\begin{figure}
 \begin{center}
  \includegraphics[scale=0.85, origin=c, angle=-90]{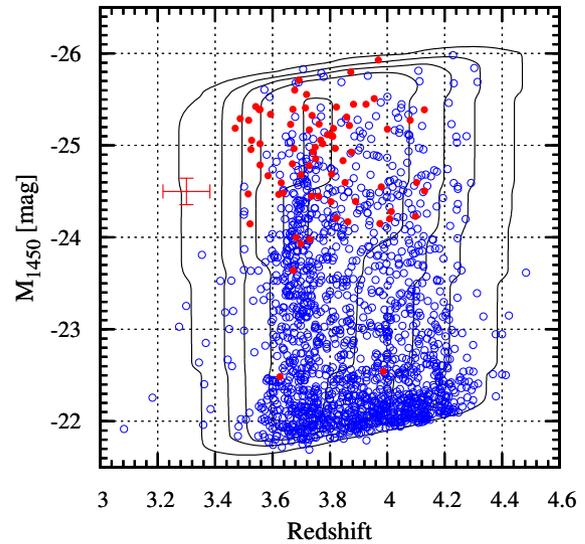}
 \end{center}
\caption{
Estimated redshift and $M_{\rm 1450}$ of the $z\sim4$ quasar candidates.
Filled red and open blue circles represent quasars with spectroscopic
and photometric redshifts, respectively. The error bar shows the
uncertainty associated with the redshift and $M_{\rm 1450}$ 
estimations. Contours represent survey area at each redshift and $M_{\rm 1450}$.
The contours are plotted at 10deg$^{2}$, 40deg$^{2}$, 70deg$^{2}$, 100deg$^{2}$, 
and 130deg$^{2}$.
\label{fig:WIDE_Red_Mabs}}
\end{figure}

We derive $M_{\rm 1450}$ either with $z_{\rm phot}$
or $z_{\rm spec}$. $M_{\rm 1450}$ is derived by
interpolating the broad-band photometry around rest-frame
1450 {\AA}. The uncertainty of $M_{\rm 1450}$ associated
with $z_{\rm phot}$ is evaluated by the difference
of $M_{\rm 1450}$ with $z_{\rm spec}$ and $z_{\rm phot}$
for the 74 objects with $z_{\rm spec}$. The resulting
$\sigma$ of the difference is 0.082 mag. The estimated $M_{\rm 1450}$ are plotted against
redshift in figure~\ref{fig:WIDE_Red_Mabs} with filled red
and open blue circles for objects with $z_{\rm spec}$ and
$z_{\rm phot}$. Most of the spectroscopically identified
objects are from the SDSS catalog, and distributed 
brighter than $-24$ mag.

\subsection{Binned quasar luminosity function at $z\sim4$}

\begin{figure}
 \begin{center}
  \includegraphics[scale=0.90]{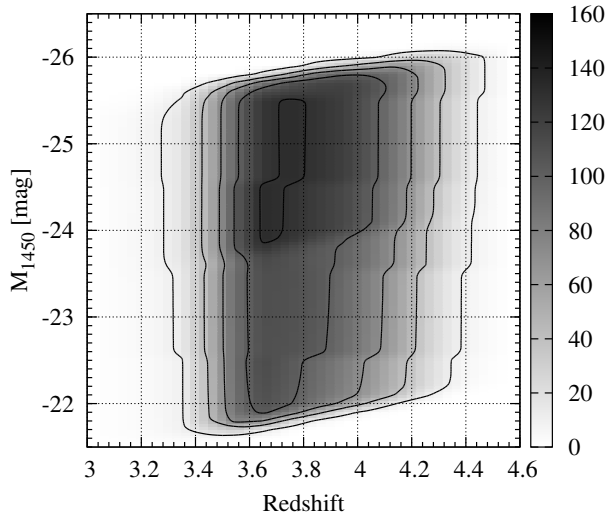}
 \end{center}
\caption{
Effective survey area as a function of redshift and $M_{\rm 1450}$.
The contours are plotted at 10deg$^{2}$, 40deg$^{2}$, 70deg$^{2}$, 
100deg$^{2}$, and 130deg$^{2}$.
The effective survey area include the efficiency of the
color selection for the mock quasars. 
The upper and lower edge is determined with $i>20.0$ and $i<24.0$
mag selection criteria. The gap seen in the middle reflects
the effect of the masks around HSC objects brighter than $i<22.0$ mag.
For objects brighter than $i<22.0$ mag, the masks are not applied.
The incompleteness of the stellarity classification is not
included in the calculation.
\label{fig:WIDE_area_z4QSOabs}}
\end{figure}

The luminosity function of quasars at $z\sim4$ is
calculated, using the photometric redshifts, $M_{\rm 1450}$
and the survey area as a function of redshift and $M_{\rm 1450}$.
We evaluate the survey area again with the library of quasar models
with templates. 
The resulting effective survey area as a function of 
redshift and $M_{\rm 1450}$
is shown with gray scale in figure~\ref{fig:WIDE_area_z4QSOabs}.
We also plot the effective survey area with contour 
in figure~\ref{fig:WIDE_Red_Mabs}. The distribution of the
sample in the redshift-$M_{\rm 1450}$ plane is consistent
with that expected from the effective survey area.
It should be noted that the effective survey area includes
the efficiency of the quasar color selection, and the
the quasar models does not contain reddened quasar SEDs (see section~\ref{sec:SEDlib}).

We estimate the binned luminosity function with 
$\sum 1/V_{a}$ estimator (e.g. \cite{Miyaji2000}) with
\begin{eqnarray}
\frac{{\rm d} \Phi \ \ \ \ \ \ }{{\rm d} M_{\rm 1450}}(M_{\rm 1450})
 = \frac{\sum_{i} V_{a} (M_{{\rm 1450},i})^{-1}}{\Delta M_{\rm 1450}}
\end{eqnarray}
where the sum is taken through objects in a $M_{\rm 1450}$ bin 
with width of $\Delta M_{\rm 1450}$. $V_{a}$ represents the available volume
for the $i$-th object with $M_{{\rm 1450},i}$ in the bin,
and is calculated as
\begin{eqnarray}
 V_{a}(M_{\rm 1450}) = \int d_{A}(z)^{2} (1+z)^{3} c \ \frac{{\rm d} \tau}{{\rm d} z}(z) A(M_{\rm 1450}, z){\rm d}z,
\end{eqnarray}
where $d_{A}(z)$ as the angular diameter distance 
and $\frac{{\rm d} \tau}{{\rm d}z}(z)$ as the look-back time per unit $z$.
$A(M_{\rm 1450},z)$ represents the survey area shown in 
figure~\ref{fig:WIDE_area_z4QSOabs}.

\begin{figure}
 \begin{center}
  \includegraphics[origin=c,angle=-90,scale=0.90]{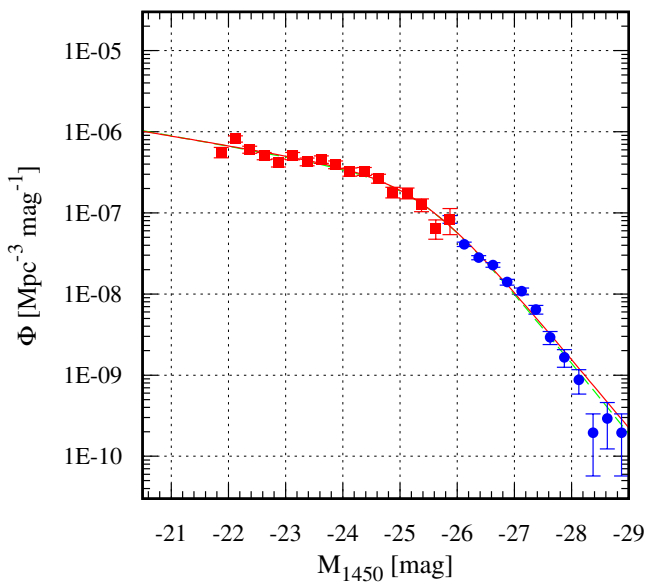}
 \end{center}
\caption{
Luminosity function of $z\sim4$ quasars derived in this work.
Red filled squares and blue filled circles represent
results based on the HSC and SDSS DR7 quasar samples, respectively.
Red solid and green dashed lines show the best-fit double power-law model with 
the maximum likelihood fit to the sample 
and the $\chi^{2}$ fit to the binned luminosity functions.
\label{fig:WIDE_z4QSO_LFnew}}
\end{figure}

The resulting luminosity function is shown in figure~\ref{fig:WIDE_z4QSO_LFnew}
with red filled squares.
Thanks to the wide and deep coverage of the HSC Wide-layer
dataset, the faint-end of the
quasar luminosity function at $z\sim4$ is
constrained with unprecedented accuracy, and the
break of the luminosity function is clearly detected
at around $M_{\rm 1450}\sim-25$ mag. The binned luminosity
function is shown in table~\ref{tab:binnedLF}. 
The uncertainty, $\sigma$, in each bin is determined
with the Poisson statistics.

\subsection{$z=4$ quasar sample from SDSS DR7}

In order to construct the luminosity function covering 
a wide luminosity range based
on quasars selected with consistent selection criteria,
we combine $z=3.6$ to $4.2$ quasars from the spectroscopically identified
quasar catalog from SDSS DR7 \citep{Schneider2010}.
We use the DR7 catalog, because the SDSS legacy survey with 
a uniform color selection for high-redshift quasars is fully
covered in the catalog. In the SDSS legacy survey, candidates
of high-redshift quasars are selected with stellar morphology
and multi-color selection for spectroscopic follow-up \citep{Richards2002}.
A uniform target selection is applied for spectroscopy,
except for the early phase of the SDSS survey.
In order to construct a statistical sample, we only consider
objects which are spectroscopically observed as a science primary
object based on the uniform target selection 
(SCIENCEPRIMARY$=1$ and TARGET selection flag in
table 2 of \citet{Schneider2010}). 
The effective survey area of the SDSS DR7 sample is
determined to be 6,248 deg$^{2}$ by \citet{Shen2012}.

% The effective survey area of the SDSS DR7 sample is estimated
% by comparing the number counts of $z=3.0$ to $5.0$ quasars
% meeting the above criteria with those of SDSS DR3 quasars
% in the same redshift range as a function of $i$ band magnitude
% (table 3 of \citet{Richards2006}). The DR7 number counts are
% 3.84 times larger than that in the DR3 sample in the magnitude
% range $i>18.2$ mag, therefore we estimate the effective survey area 
% of the DR7 sample as 6,228 deg$^{2}$
% (effective survey area of DR3 sample, 1,622 deg$^{2}$
% \citep{Richards2006}, multiplied by 3.84).
%
The selection function of the uniform target selection is
evaluated in \citet{Richards2006} as a function of redshift and
magnitude. In the redshift range between $z=3.6$ and $4.2$,
the SDSS target selection efficiency is estimated to be 
larger than 90\%, mostly $\sim$100\% (figure 6 of \citet{Richards2006}).
The target selection for high-redshift quasar is complete
down to $i<20.2$ mag, therefore we select 1,260 quasars brighter
than $M_{\rm i}(z=2) < (z-3.0) - 26.85$ as a statistically
complete sample in the above effective area (figure 17 of \citet{Richards2006}).
We assume that the sample is complete in the redshift range
above the absolute magnitude limit. 
Utilizing the absolute magnitude limit as a function of redshift, 
we calculate effective survey volume for each object with their $M_{\rm i}(z=2)$.
UV absolute magnitudes, $M_{\rm 1450}$, of the quasars in the sample
are derived from the broad-band SDSS photometry of the
quasars in the same way for the HSC $z=4$ quasars.

The resulting luminosity function is shown in figure~\ref{fig:WIDE_z4QSO_LFnew}
with blue filled circles and summarized in table~\ref{tab:binnedLF}. 
The second column indicates the number of the quasars in each absolute 
magnitude bin after correcting for the contamination rate.
The luminosity function smoothly connects
to the luminosity function determined with the HSC $z=4$ quasar candidates.
Because the magnitude range of the HSC $z=4$ quasar candidates
is fainter than $i>20.0$ mag and the SDSS DR7 sample is $i<20.2$ mag,
there is no overlap in the bin of the luminosity function.

\begin{table}
  \caption{Binned $z=4$ quasar luminosity function}\label{tab:binnedLF}
  \begin{center}
    \begin{tabular}{crccr}
\hline
$M_{\rm 1450}$ & $N_{\rm corr}$ & $\log (\Phi)$\footnotemark[$\dag$]  & $\sigma$\footnotemark[$*$] \\ 
    (mag)      &                &                                     &                            \\ \hline
\multicolumn{4}{c}{HSC S16A-Wide2} \\ \hline
$-21.875$ &  47.7 & $-6.253$ & 80.733 \\ 
$-22.125$ & 129.0 & $-6.088$ & 71.892 \\ 
$-22.375$ & 103.7 & $-6.219$ & 59.367 \\ 
$-22.625$ &  92.8 & $-6.298$ & 52.208 \\ 
$-22.875$ &  78.5 & $-6.382$ & 46.855 \\ 
$-23.125$ &  95.8 & $-6.297$ & 51.538 \\ 
$-23.375$ &  81.4 & $-6.369$ & 47.366 \\ 
$-23.625$ &  90.4 & $-6.341$ & 47.960 \\ 
$-23.875$ &  84.7 & $-6.405$ & 42.783 \\ 
$-24.125$ &  74.0 & $-6.493$ & 37.343 \\ 
$-24.375$ &  76.0 & $-6.487$ & 37.385 \\ 
$-24.625$ &  63.0 & $-6.577$ & 33.384 \\ 
$-24.875$ &  43.0 & $-6.745$ & 27.422 \\ 
$-25.125$ &  42.0 & $-6.756$ & 27.088 \\ 
$-25.375$ &  30.0 & $-6.899$ & 23.032 \\ 
$-25.625$ &  14.0 & $-7.189$ & 17.287 \\ 
$-25.875$ &   8.0 & $-7.077$ & 29.581 \\ 
\hline 
\multicolumn{4}{c}{SDSS DR7 $z=3.6-4.2$} \\ \hline
$-26.125$ & 260.0 & $-7.387$ & 2.545 \\ 
$-26.375$ & 287.0 & $-7.552$ & 1.657 \\ 
$-26.625$ & 234.0 & $-7.642$ & 1.490 \\ 
$-26.875$ & 144.0 & $-7.853$ & 1.169 \\ 
$-27.125$ & 111.0 & $-7.966$ & 1.026 \\ 
$-27.375$ &  66.0 & $-8.192$ & 0.791 \\ 
$-27.625$ &  30.0 & $-8.534$ & 0.534 \\ 
$-27.875$ &  17.0 & $-8.781$ & 0.402 \\ 
$-28.125$ &   9.0 & $-9.057$ & 0.292 \\ 
$-28.375$ &   2.0 & $-9.710$ & 0.138 \\ 
$-28.625$ &   3.0 & $-9.534$ & 0.169 \\ 
$-28.875$ &   2.0 & $-9.710$ & 0.138 \\ 
\hline
    \end{tabular}
  \end{center}
\begin{tabnote}
\footnotemark[$\dag$] In unit of Mpc$^{-3}$ mag$^{-1}$.
\footnotemark[$*$] In unit of $1.0\time10^{-9}$Mpc$^{-3}$ mag$^{-1}$.
\end{tabnote}
\end{table}

\subsection{Double power-law model}

The quasar luminosity function is generally well described
by a double power-law form with
\begin{eqnarray}
\phi (M_{\rm 1450}, z) & \nonumber \\
= & \frac{\phi^{*}}{10^{0.4(\alpha+1)(M_{\rm 1450}-M_{*})}+10^{0.4(\beta+1)(M_{\rm 1450}-M_{*})}}
\end{eqnarray}
where $M_{*}$ is the absolute magnitude of the knee and $\phi^{*}$ is the number
density at that luminosity. $\alpha$ and $\beta$ are power-law slopes of
the faint- and bright-ends, respectively.
The double power-law model is fitted to the $z=4$ quasar samples from 
the HSC and SDSS DR7 with the maximum likelihood method \citep{Marshall1983}. 

We use 
\begin{eqnarray}
 \mathcal{L} = && -2 \sum_{i}^{N_{\rm obj}} \ln \left[ \frac{N(M_{{\rm 1450},i}, z_{i})}{\iint N(M_{\rm 1450}, z) {\rm d}M_{\rm 1450} {\rm d}z} \right],
\end{eqnarray}
which is modified version of the maximum likelihood estimator for 
luminosity function (e.g. \cite{Miyaji2000}) 
$N(M_{\rm 1450}, z)$ is the expected number of object with a model per unit absolute magnitude and redshift inverval,
\begin{eqnarray}
	N(M_{\rm 1450}, z) = \frac{{\rm d}\Phi^{\rm model}}{{\rm d}M_{\rm 1450}} d_{A}(z)^{2} (1+z)^{3} c \ \frac{{\rm d}\tau}{{\rm d}z}(z) A(M_{\rm 1450}, z).
\end{eqnarray}
The normalization of the best-fit model is not constrained with the likelihood estimator,
we determine the normalization such that the expected number of object from the model
matches $N_{\rm obj}$. Uncertainties are evaluated by fixing a parameter to a value around
its best-fit value and minimizing $\mathcal{L}$ with the other parameters. 
One sigma uncertainty of the parameter is determined by the range whose 
minimum $\mathcal{L}$ is larger by less than 1 from the best-fit $\mathcal{L}$.
The uncertainty associated with the normalization is determined by the Poisson statistics.
The best-fit parameters and associated uncertainties are summarized in the first line of table~\ref{tab:fitting}.
The best-fit model is shown with red solid line in figure~\ref{fig:WIDE_z4QSO_LFnew}.
The red solid line matches well with the binned luminosity
function.

We also fit the binned luminosity function with a double power-law model
through the $\chi^{2}$ minimization. The resulting best fit parameters
are summarized in the second line of table~\ref{tab:fitting} and the best-fit
function is shown in figure~\ref{fig:WIDE_z4QSO_LFnew} as green dashed line.
The best-fit parameters are consistent with each other.

\begin{table*}
  \caption{Parameters of the best fit $z=4$ quasar luminosity function with a double power-law model}\label{tab:fitting}
  \begin{center}
    \begin{tabular}{lccccl}
\hline
	    \multicolumn{1}{c}{Method} & $\log(\Phi^{*})$\footnotemark[$\dag$] & $\alpha$    & $\beta$      & $M_{\rm 1450}^{*}$ & \multicolumn{1}{c}{Note} \\
                           &                          & [Faint End]     & [Bright End]   &    (mag)        &    \\ \hline
 Maximum Likelihood        & $2.66\pm0.05$            &  $-1.30\pm0.05$ & $-3.11\pm0.07$ & $-25.36\pm0.13$ & Fitting to the sample \\
 $\chi^{2}$ minimization   & $2.41\pm0.56$            &  $-1.32\pm0.08$ & $-3.18\pm0.11$ & $-25.44\pm0.19$ & Fitting to the binned luminosity function \\
\hline
    \end{tabular}
  \end{center}
\begin{tabnote}
\footnotemark[$\dag$] In unit of $1\times10^{-7}$ Mpc$^{-3}$ mag$^{-1}$.
\end{tabnote}
\end{table*}

\section{Discussion}

\subsection{Comparison with previous results on the $z=4$ quasar luminosity function} 

\begin{figure}
 \begin{center}
  \includegraphics[origin=c,angle=-90,scale=0.90]{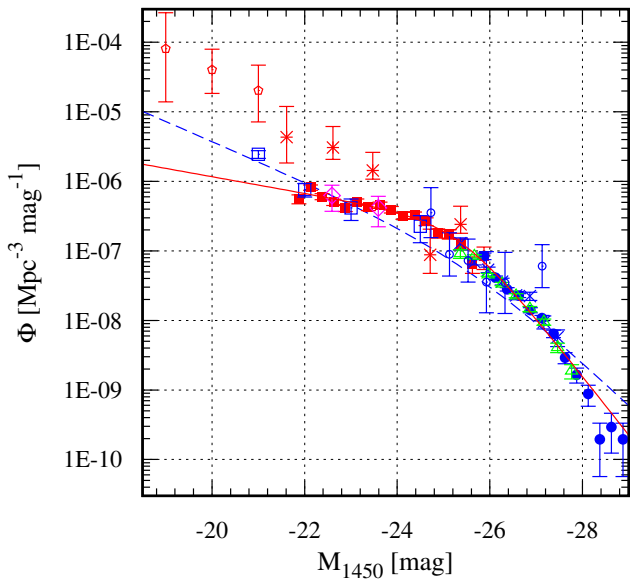}
 \end{center}
\caption{
Luminosity function of $z\sim4$ quasars derived in this work
(red filled squares:HSC and blue filled circles:SDSS DR7) 
compared with previous results 
(blue crosses: 
$3.5<z<4.0$ luminosity function from 
SDSS DR3 \citep{Richards2006},
green open triangles:
$3.5<z<4.0$ luminosity function with
SDSS DR7 quasar sample \citep{Shen2012},
blue open circles: 
$3.5<z<4.0$ luminosity function derived with time-variability
selection (\cite{Palanque2013}),
pink open diamonds:
$3.7<z<4.7$ luminosity function derived from the COSMOS
quasar survey (\cite{Niida2016}),
blue open squares:
$3.5<z<5.0$ luminosity function derived from the COSMOS
quasar survey (\cite{Masters2012}),
red asterisks:
$3.7<z<5.1$ luminosity function derived from NDWFS
(\cite{Glikman2011}). For the latter three luminosity functions,
their normalizations are corrected by factor 1.2 to 
correct for the difference in the covered redshifts.
Red pentagons show $z=4.0-4.5$ X-ray-selected AGN 
luminosity function from \citet{Giallongo2015}.
The normalization is corrected by factor 1.6. 
Blue dashed line is the best-fit luminosity function 
from the COSMOS survey (\cite{Masters2012}) after correcting
for the normalization.
\label{fig:WIDE_z4QSO_LF}}
\end{figure}

The luminosity function derived in this work is compared
with previous work in figure~\ref{fig:WIDE_z4QSO_LF}.
In the bright-end, the binned luminosity function of
$z=4$ SDSS DR7 quasar is fully consistent with those at $z=3.75$ with
the DR3 sample plotted with blue crosses \citep{Richards2006}
and the DR7 sample plotted with green open triangles \citep{Shen2012}.
We convert $M_{i}(z=2)$ used in those papers to $M_{\rm 1450}$
with $M_{\rm 1450}=M_{i}(z=2) + 1.486$ \citep{Richards2006}.
Thanks to the larger effective area of the DR7 sample, 
the luminosity function from the DR7 sample extends to
higher luminosities than the DR3 sample. 

Around the knee of the luminosity function, 
\citet{Palanque2013} evaluate the luminosity function
based on a quasar sample selected by a variability selection.
The luminosity function is described with $M_{g}(z=2)$
and we convert $M_{\rm 1450}=M_{g}(z=2) + 1.272$
based on figure 12 of \citet{Palanque2013}.
The binned luminosity function covering $3.5<z<4$ is plotted
with blue open circles in the figure. The binned luminosity
function is fully consistent with the luminosity function 
of this work within the uncertainty.

In \citet{Ikeda2011}, \citet{Glikman2011}, \citet{Masters2012}, and \citet{Niida2016},
the quasar luminosity functions are derived down to $M_{\rm 1450}=-21$ mag
in deeper but narrower surveys than the SDSS.
These functions are determined in the redshift range
$3.7<z<4.7$, $3.7<z<5.1$, $3.5<z<5.0$, and $3.7<z<4.7$ respectively. They
cover a higher redshift range with the average 
redshift of the samples of $z=4.0$, 
therefore we compare with them after correcting for the evolution 
effect. In this redshift range, the number density of
quasars evolves as $(1+z)^{-6.9}$ \citep{Richards2006}, 
and we correct for the difference by multiplying 1.2 with the 
averages of the samples. 
The binned luminosity functions in the results are shown 
with pink open diamonds, red crosses, and blue open squares
for \citet{Niida2016}, \citet{Glikman2011}, and \citet{Masters2012}
samples, respectively, in figure~\ref{fig:WIDE_z4QSO_LF}. 
The binned luminosity functions derived in the COSMOS regions
(\cite{Niida2016}; \cite{Masters2012}) matches well that
of this work, except for the faintest luminosity bin at $M_{\rm 1450}=-21$ mag.
They select $z=4$ quasars with stellar morphology and
dropout selection method in the similar way to this study. 
On the other hand, the luminosity function by \citet{Glikman2011}
based on NOAO Deep Wide-Field Survey (NDWFS) dataset
shows significantly higher number density than the other
results. They use similar selection method with dropout
color and stellar morphology, but they use loose selection
criteria for stellarity based on ground-based imaging data
with FWHM of \timeform{1"},
and it would be possible that their sample
is significantly contaminated by extended non-AGN objects 
below $R>22.5$ mag (figure 4 in \cite{Glikman2010}).
It should be noted that the number density of LBGs rapidly
increases below $i>23$ mag, and they can severely contaminate
with the loose stellarity selection as discussed
in section~\ref{sec:cont_extend}.

Although the binned luminosity functions in the COSMOS
regions are consistent with this work, the best-fit
luminosity function in \citet{Masters2012} has much
steeper slope than that from this work. The best-fit
luminosity function is shown with blue dashed line after 
correcting for the redshift difference by multiplying 1.2. The steeper slope
is mostly caused by the excess number counts at $M_{\rm 1450}=-21$ mag,
and cannot reproduce the luminosity function brighter than the
knee. In section~\ref{sec:XLF}, we further discuss the difference
in the best-fit luminosity functions at $z=4$.

By integrating the best-fit luminosity function,
we estimate the UV luminosity density at 1450 {\AA} of the quasars at $z=4$
to be $\epsilon_{\rm 1450}=3.2\times10^{24}$ erg s$^{-1}$ Hz$^{-1}$ Mpc$^{-3}$, which
is similar to that with the best-fit luminosity function by \citet{Masters2012}
($\epsilon_{\rm 1450}$ = $3.1\times10^{24}$ erg s$^{-1}$ Hz$^{-1}$ Mpc$^{-3}$
after correcting for the evolution factor of 1.2).
\citet{Giallongo2015} derive the UV luminosity function of X-ray selected
AGNs and argue that the AGN emissivity of UV ionizing photons could be as high
as the value required to keep the intergalactic medium highly ionized. 
Their UV luminosity function is shown with red pentagons in figure~\ref{fig:WIDE_z4QSO_LF}.
We correct for the redshift evolution by multiplying 1.6 to their number density
assuming the middle point of their redshift coverage ($z=4.0-4.5$) and 
the number density evolution with $(1+z)^{-6.9}$. 
Their number density is more than one order of magnitude higher than the
extrapolation of our best-fit luminosity function at $z=4$.
Their estimated UV luminosity density is $18.3\times10^{24}$ erg s$^{-1}$ Hz$^{-1}$ Mpc$^{-3}$
after correcting for the evolution factor. Our estimate is about 6 times smaller than
their value, and suggests that optically-selected stellar blue quasars are not 
the main contributor to the cosmic reionization.

\subsection{Evolution of the quasar luminosity function in the early universe}
\label{sec:LFevol}

\begin{figure*}
 \begin{center}
  \includegraphics[angle=-90,scale=0.85]{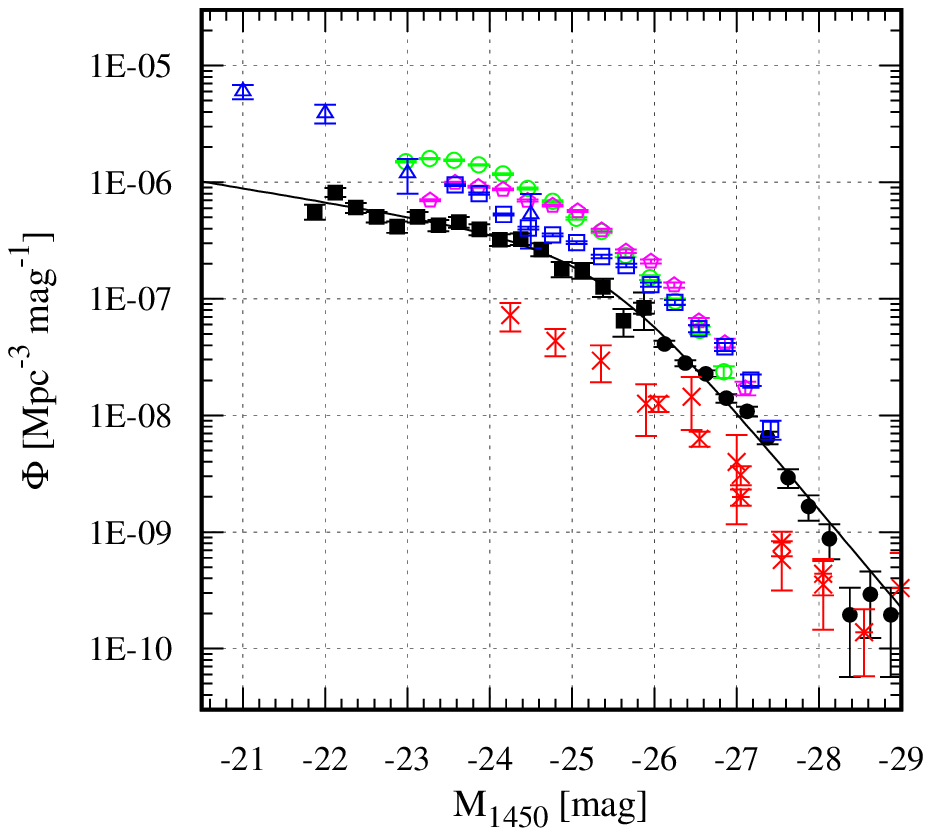}
  \includegraphics[angle=-90,scale=0.85]{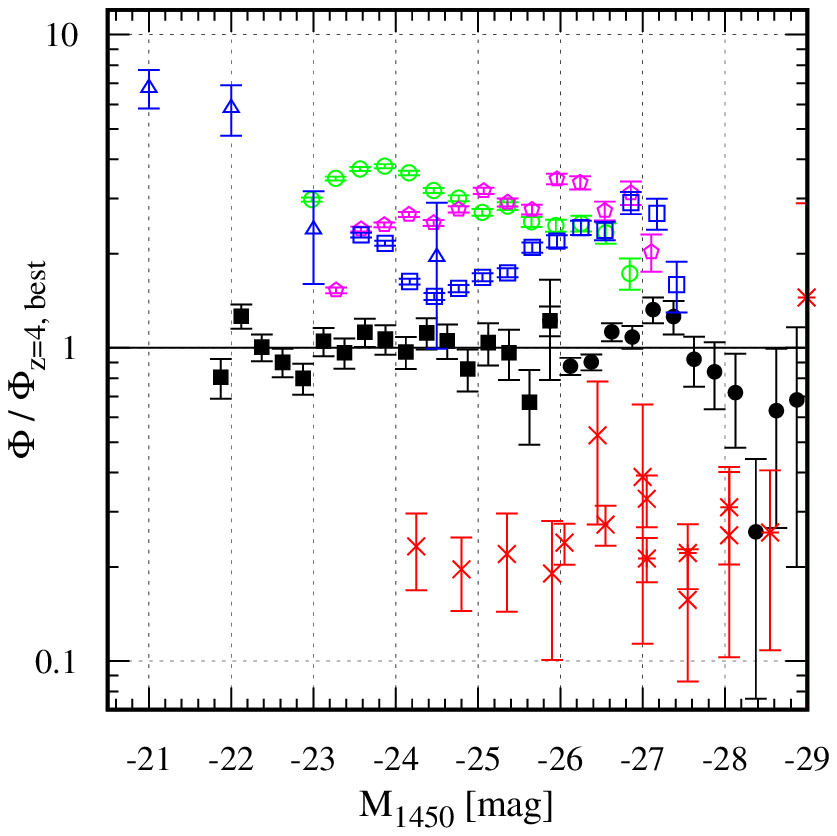}
 \end{center}
\caption{
Luminosity functions of quasars at $2<z<5$. 
Black filled squares and circles represent $z=4$ quasar
luminosity function from the HSC and SDSS DR7 samples, respectively.
Solid line indicates the best-fit double power-law model to the $z=4$
luminosity function.
Quasar luminosity functions at
$z=5$ (Red crosses and asterisks; \cite{McGreer2013} and \cite{Yang2016}, respectively), 
$z=3.2$ from SDSS/BOSS (Blue open squares; \cite{Ross2013})
and COSMOS (Blue open triangles; \cite{Masters2012}),
$z=2.7$ (Pink open pentagons; \cite{Ross2013}),
and $z=2.3$ (Green open circles; \cite{Ross2013}) are
shown. Binned luminosity functions with $M_{i}(z=2)$ are
converted to $M_{\rm 1450}$ with $M_{\rm 1450}=M_{i}(z=2)+1.486$
\citep{Richards2006}.
Left) Binned luminosity function. Right) 
Number ratio of the binned luminosity function at each redshift
to the best-fit model at $z=4$.
\label{fig:WIDE_z4QSO_LFcomp}}
\end{figure*}

\begin{figure*}
 \begin{center}
  \includegraphics[angle=-90,scale=0.95]{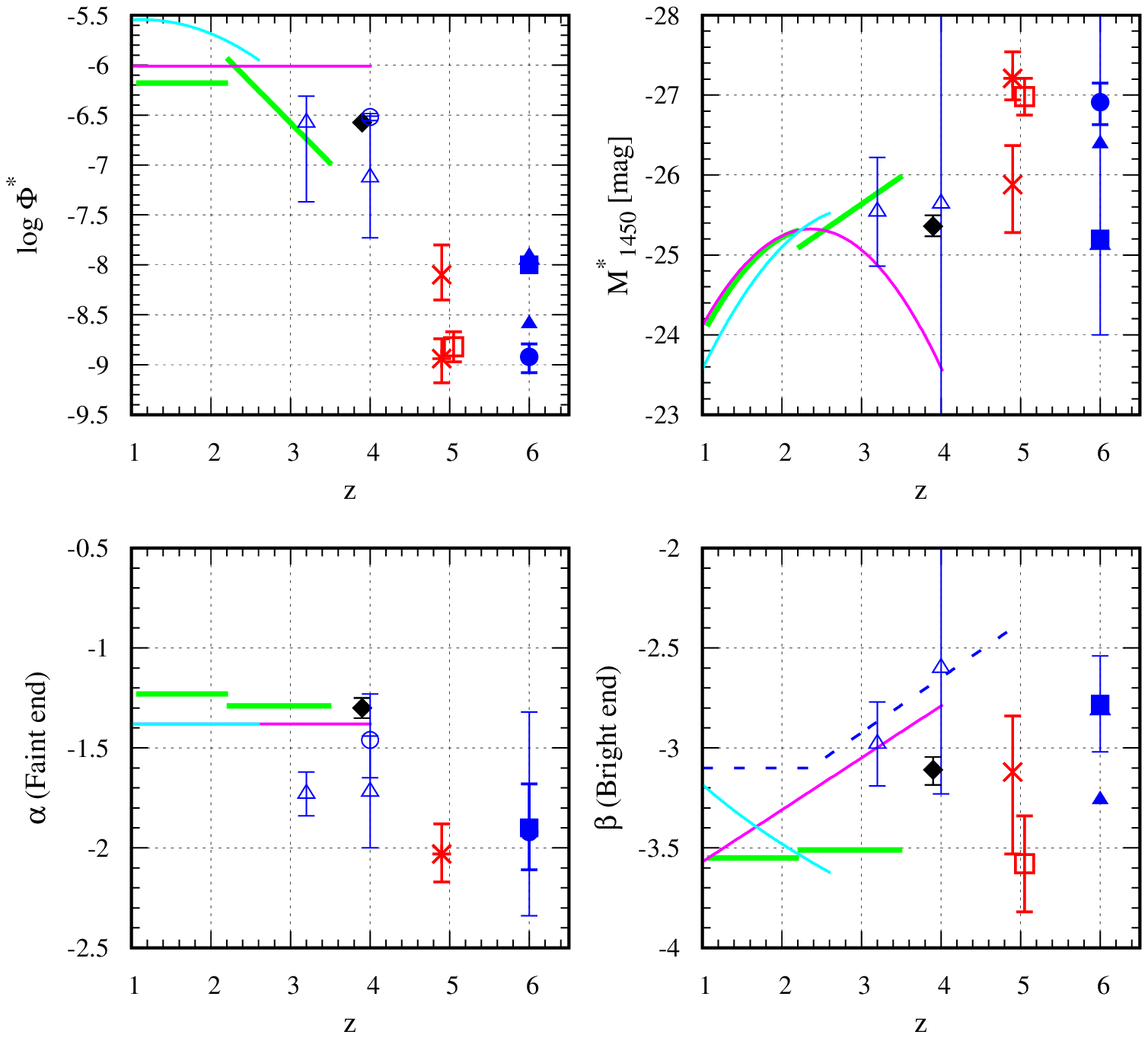}
 \end{center}
\caption{
Best-fit quasar luminosity function parameters (filled diamonds) 
are compared with those in literature.
Green solid lines; best-fit PLE and LEDE luminosity function
parameters from \citet{Ross2013} below and above $z=2.2$,
respectively, 
cyan solid lines; best-fit LEDE luminosity function
parameters from \citet{Croom2009},
pink solid line; $\beta$ varying luminosity evolution 
fit from \citet{Bongiorno2007},
blue dashed line; variable power-law model for bright-end
luminosity function from \citet{Richards2006},
blue open triangles; double power-law fit to
$z\sim3.2$ and $z\sim4$ luminosity functions from 
\citet{Masters2012}, 
blue open circles; double power-law fit to 
$z\sim4$ luminosity function from \citet{Niida2016}, 
red cross and asterisk; fit to $z=5$ luminosity function 
from \citet{McGreer2013} with fixed $\alpha$ and
$\beta$, respectively,
red open square; $z=5$ luminosity function from \citet{Yang2016}, 
blue filled triangles; bright-end $z\sim6$ luminosity function 
from \citet{Willott2009} with fixed $\alpha=-1.5$ (large)
and $\alpha=-1.8$ (small),
blue filled circle; $z\sim6$ luminosity function
from case 1 of \citet{Kashikawa2015},
blue filled square with thin errorbar; $z\sim6$ luminosity
function from \citet{Jiang2016}.
\label{fig:z4QSOLF_param}}
\end{figure*}

The evolutionary trend in the luminosity function of quasars above the peak
of their number density at $z\sim2-3$ is a fundamental observable to understand
the early growth of SMBHs. In the left panel of figure~\ref{fig:WIDE_z4QSO_LFcomp},
the binned quasar luminosity function at $z=4$ is compared with those at
$z\sim2.3$ (green open circles; \cite{Ross2013}), 
$z\sim2.7$ (pink open pentagons; \cite{Ross2013}),
$z\sim3.2$ (blue open squares; \cite{Ross2013}; blue open triangles; \cite{Masters2012}),
and $z\sim5$ (red crosses; \cite{McGreer2013}). 
In the right panel of figure~\ref{fig:WIDE_z4QSO_LFcomp}, 
the ratios of the binned quasar luminosity functions to the best-fit 
double power-law luminosity function at $z=4$ are shown.
In figure~\ref{fig:z4QSOLF_param}, we also plot the best-fit double power-law
model parameters as a function of redshift. If multiple fitting results
with different evolutionary scenarios, for example pure-luminosity and 
pure-density evolution scenarios, are provided in a paper,
we pick up the model parameters that reproduce the binned luminosity 
function better.

The slopes of the bright-end of the
luminosity functions distribute around $\beta=3$ above $z=3$, 
except for the Luminosity Evolution and Density Evolution model in 
\citet{Ross2013} and \citet{Yang2016} with steeper bright-end slopes. 
The value in \citet{Ross2013} is
closer to those observed at $z<2$ (e.g. \cite{Croom2009}) and could be affected by quasars
in the redshift range around $z=2$. If we compare their binned luminosity functions
at $z=2.3$ and $z=2.7$ with that at $z=4$, the slopes above the knee
seem to be similar to each other except for the most luminous bins. 
The systematically flatter bright-end slope above $z=3$ than that at $z<2$ 
is consistent with the trend reported in \citet{Richards2006}, \citet{Bongiorno2007},
and \citet{Richards2009}.
The constant bright-end slope above $z=3$
would suggest that we do not need a luminosity dependent evolution model
above the knee to explain the increase of the number density
at a fixed luminosity \citep{Steffen2006}.

The evolution of the faint-end shows a different trend. 
The luminosity functions at $z=2.3$ and $2.7$ 
have a similar shape to the $z=4$ luminosity function and 
the best-fit faint-end slope at $z=2.2-3.5$ is 
$\alpha_{\rm z=2.2-3.5}=-1.29^{+0.15}_{-0.03}$ \citep{Ross2013},
which is consistent with the best-fit value at $z=4$ within the 1$\sigma$
uncertainty. Flat faint-end slope is also supported by 
\citet{Bongiorno2007}, \citet{Fontanot2007}, and \citet{Niida2016}. 
On the other hand, the best-fit model at $z=3.2$ has
$\alpha_{\rm z=3.2}=-1.73\pm0.11$ \citep{Masters2012}, which is much
steeper than that at $z=4$. Such difference can be seen in the shape
of the luminosity functions, the binned $z=3.2$ luminosity function does not show
clear knee in it. The faint-end slopes at $z=5$ and $6$ could be
steeper than that at $z=4$ (\cite{McGreer2013}; \cite{Kashikawa2015}).
However, the current quasar luminosity function at $z=5.0$ 
could not be deep enough to clearly detect the knee of
the luminosity function, although the best-fit value for the faint-end
would imply a steeper slope, $\alpha_{\rm z=5}=-2.03^{+0.15}_{-0.14}$
\citep{McGreer2013}. Furthermore, the constraint on the faint-end of the
quasar luminosity function at $z=6.0$ is based on a few quasars
\citep{Kashikawa2015}. 

The overall evolutionary trend from $z=2$ to $4$ 
does not support the higher break luminosities 
and steeper faint-end slopes with redshifts suggested above $z=5$ (\cite{McGreer2013}; \cite{Kashikawa2015}). 
In the right panel of figure~\ref{fig:WIDE_z4QSO_LFcomp}, 
luminosity functions at $z=2.3$, $2.7$, and $5$ show relatively flat distribution of the
ratio, which would imply pure density evolution of the quasar luminosity function.
Such trend is consistent with the evolutionary trend seen in the X-ray luminosity function 
at $z>3$ \citep{Vito2014}.
The shape of the luminosity function at $z=3.2$ shows
a different evolutionary trend. 
The number ratio increases toward the high-luminosity and low-luminosity
ends and the ratio is small in the mid-luminosity range around the knee.

\subsection{Comparison with the X-ray AGN luminosity function}
\label{sec:XLF}

Comparing the luminosity function of the optically-selected
quasars with that of X-ray-selected AGNs in the same redshift range, 
we can infer the fraction of blue stellar quasars among the 
X-ray AGN population, which includes heavily-obscured AGNs. 
Recent analysis of samples of X-ray selected AGNs indicates that
the fraction of heavily-obscured AGNs is higher at $z>3$
than that in the local universe (\cite{Hiroi2012}; Vito et al.\ \yearcite{Vito2013}, \yearcite{Vito2014}), 
and it is possible that the optical selections are missing
such heavily-obscured AGNs. In this
discussion, we assume both of the AGN populations have the 
same overall SED, but the optical selection can be affected by 
the dust extinction and host galaxy contamination.

We convert the $z=4$ quasar $M_{\rm 1450}$ luminosity function 
to the hard X-ray 2--10~keV luminosity function based on 
the relation between the monochromatic luminosity at 2500 {\AA}, 
$l_{\rm 2500{\AA}}$, and 2~keV, $l_{\rm 2keV}$ for non-obscured broad-line AGNs
(e.g. \cite{Vignali2003}; \cite{Strateva2005}; \cite{Steffen2006};
\cite{Young2010}).
$l_{\rm 2500{\AA}}$ is calculated from $M_{\rm 1450}$ assuming
a power-law spectrum with index of $\alpha=-0.46$.
Then, we apply the $l_{\rm 2500{\AA}}$ - $l_{\rm 2keV}$ 
relation in \citet{Steffen2006}, which is derived from 333 optically-selected AGNs, as
\begin{equation}
\log (l_{\rm 2keV}) = 0.721 \log(l_{\rm 2500{\AA}}) + 4.531.
\end{equation}
Finally, we convert $l_{\rm 2keV}$ to $L_{\rm 2-10~keV}$ 
assuming the typical X-ray spectrum of broad-line quasars, 
a power-law spectrum with a photon index of $\Gamma=1.8$.

\begin{figure*}
 \begin{center}
  \includegraphics[angle=-90,scale=0.85]{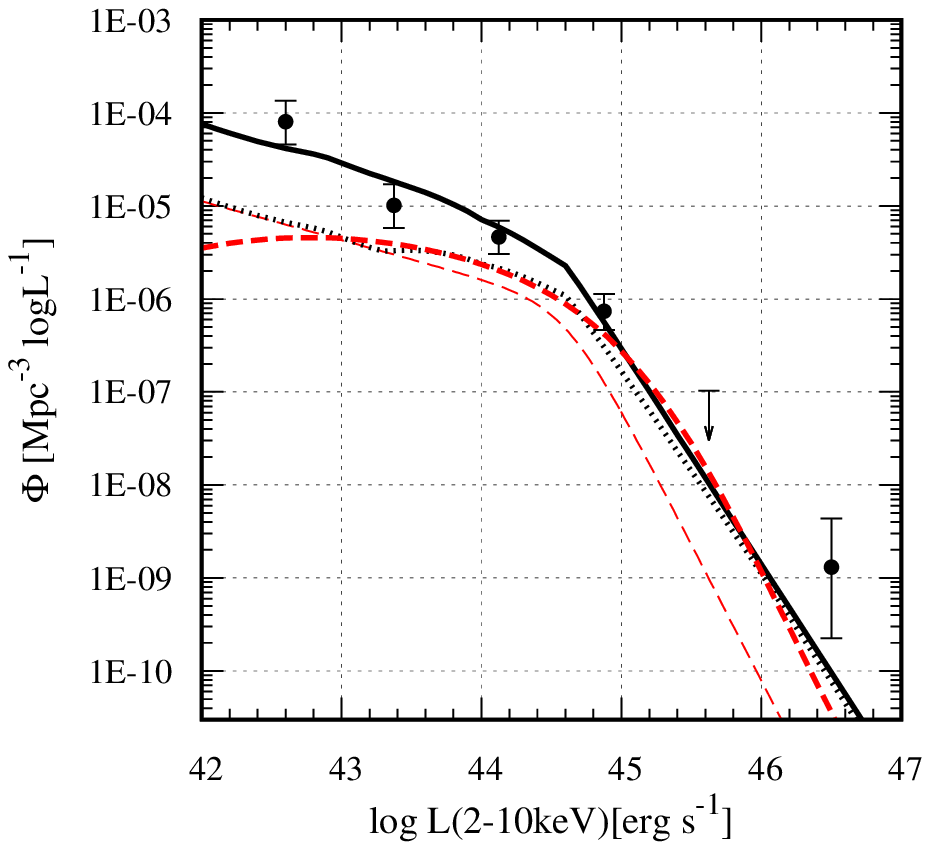}
  \includegraphics[angle=-90,scale=0.85]{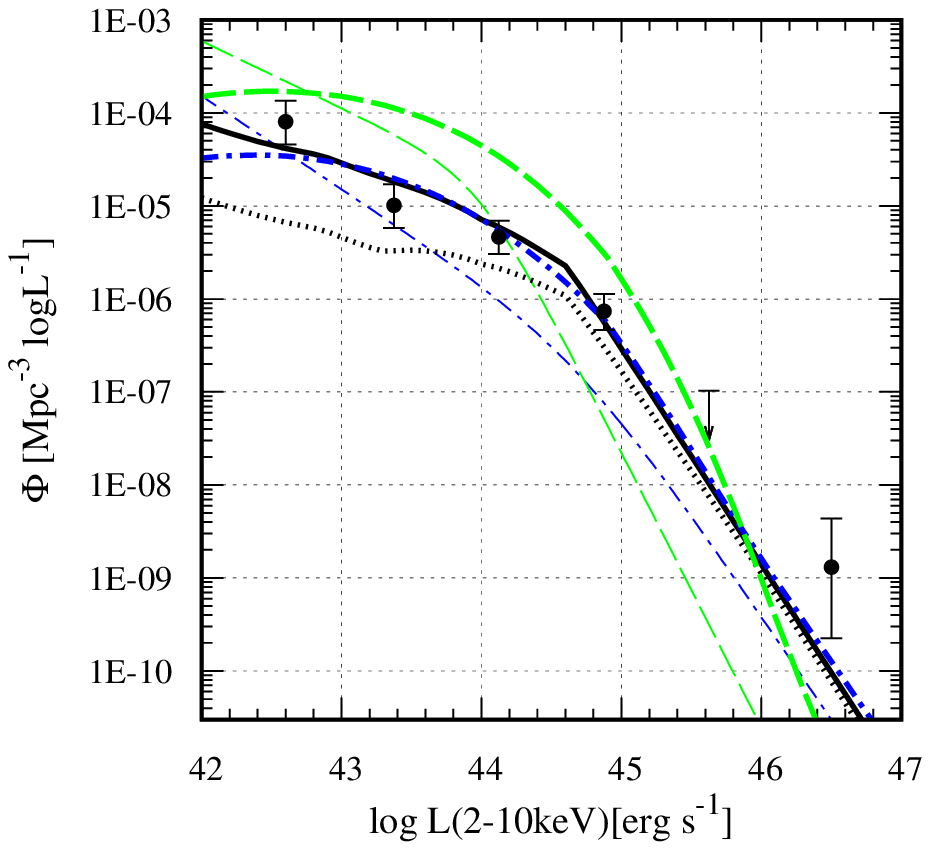}
 \end{center}
\caption{
Left) $z=4$ quasar 2--10~keV luminosity function predicted from the
UV luminosity function of the HSC and SDSS DR7 samples
(red dashed lines). Thick and thin red dashed lines indicate 
the results with and without the effect of the scatter in 
the $l_{\rm 2500}$ vs. $l_{\rm 2keV}$ relation.
Thick solid line is the best-fit LDDE model of the
2--10~keV X-ray luminosity function of
X-ray selected AGNs at $z=3.9$ \citep{Ueda2014}. 
Data points of the X-ray luminosity function derived with
the ratio of the number of the sample to that of the model prediction
are plotted with the filled circles with $1 \sigma$ error bar. 
The upper limit without point indicate 90\% upper limit.
Right) 2--10~keV luminosity functions based on the best-fit UV
luminosity functions from \citet{Masters2012} (blue dot-dashed lines) 
and \citet{Giallongo2015} (green long-dahsed lines). 
Thick and thin lines represent the results with and without
the effect of the scatter. Black lines are the same as in the left panel.
\label{fig:WIDE_z4QSO_XLF}}
\end{figure*}

The resulting hard X-ray luminosity function derived from the optical quasar
luminosity function is shown in the left panel of figure~\ref{fig:WIDE_z4QSO_XLF} as
red thin dashed line. The thick solid line indicates the hard X-ray 
luminosity function of Compton-thin AGNs at $z=3.9$ based on the best-fit
luminosity-dependent density evolution (LDDE) model by \citet{Ueda2014}.
The luminosity function of X-ray-selected AGNs outnumber in the 
entire luminosity range compared to that of the optically-selected quasars, and the
discrepancy is broadly consistent with recent comparisons
\citep{Marchesi2016}. However, we need to note that the simple
one-to-one conversion from UV to X-ray luminosity (\cite{Masters2012}; \cite{Marchesi2016})
may not be appropriate to calculate the converted luminosity function.
As discussed in \citet{Steffen2006}, at a fixed $l_{\rm 2500{\AA}}$,
the $l_{\rm 2keV}$ values show significant scatter around the mean 
value, and such scatter can broaden the luminosity
function towards the bright-end due to the steep slope of the luminosity function.
Therefore, we convert the UV luminosity function by assuming that the UV luminosity is the primary parameter
of the quasar luminosity and that the X-ray luminosity have a scatter
around the above relation.
Using the rms scatter of $\log (l_{\rm 2keV})$
measured in each $l_{\rm 2500{\AA}}$ bin (table 5 of \citet{Steffen2006}), 
we derive the rms scatter as a function of 
$\log (l_{\rm 2500{\AA}})$. 

The resulting hard X-ray luminosity function including the scatter is
shown by the red thick dashed line in the left panel of figure~\ref{fig:WIDE_z4QSO_XLF}.
We consider the luminosity range above $M_{\rm 1450}<-16.0$ mag, which
corresponds to $\log L_{\rm 2-10~keV}({\rm erg} \ {\rm s}^{-1})=41.91$ for
the above conversions. Due to the minimum luminosity, the faint-end of
the converted luminosity function shows decline toward fainter luminosity.
The resulting X-ray luminosity function is consistent with the best-fit
LDDE model of the X-ray luminosity function at $z=3.9$ above the knee of the luminosity
function. The consistency suggests that the current selection of the $z\sim4$ quasars
does not miss a significant fraction of quasars that are affected by heavy-obscuration.

Below the knee of the luminosity function, the number density of
the converted luminosity function shows a deficiency compared to the hard X-ray luminosity 
function. The deficiency can caused by the contribution of obscured and/or
less-luminous AGNs which are missed in our selection of $z\sim4$ quasars
based on stellar morphology and blue UV color. Obscured AGNs and less-luminous
AGNs, i.e. Seyfert galaxies and low-luminosity AGNs, can have extended morphology
dominated by their host galaxies and/or red UV color due to the dust extinction.
In the left panel of figure~\ref{fig:WIDE_z4QSO_XLF}, the X-ray AGN luminosity function for 
the less-absorbed AGNs with $\log N_{\rm H}({\rm cm}^{-2})<22$ is shown with thick dotted line. 
We assume the column density distribution derived in \citet{Ueda2014}.
The luminosity function of less-absorbed AGNs matches well with the converted
luminosity function of the optically-selected quasars down to 
$\log L_{\rm 2-10keV}$ (erg \ s$^{-1}$)$=43$, supporting the deficiency below
the knee is caused by the contribution from the obscured AGNs.

The consistency between the luminosity functions of optically-selected
quasars and X-ray-selected AGNs above the knee could show different luminosity dependence of
obscured fraction at $z>3$ from that determined with the X-ray-selected AGNs (\cite{Vito2014}).
\citet{Vito2014} show that the fraction of obscured AGN with the hydrogen column density,
$\log N_{\rm H}$ (cm$^{-2}$), larger than 23, does not strongly depend on 
luminosity in the luminosity range between $\log L_{\rm 2-10keV}$ (erg s$^{-1}$)$=43$
and 45 with $0.54\pm0.05$. The upper luminosity corresponds to $M_{\rm 1450}=-26.7$ mag
and the luminosity function does not show significant deficiency of optically-selected
quasars in the luminosity range. This discrepancy would be explained with the
luminosity dependence of the relation between broad-line AGN fraction and
hydrogen column density measured in X-ray \citep{Ueda2003};
large hydrogen column density can be observed among luminous blue broad-line quasars
(e.g. \cite{Akiyama2000}; \cite{Merloni2014}).

We need to note that the above conversion depends on the assumed 
$l_{\rm 2500{\AA}}$-$l_{\rm 2keV}$ relation. If we apply the relation derived
in \citet{Young2010}, the converted luminosity functions with 
their results based on CENSORED and SPECTRA samples show smaller and larger
number density than the X-ray luminosity function above the knee, respectively.

We also apply the same conversion to the best-fit model of the $z=4$ luminosity
function from \citet{Masters2012} and from \citet{Giallongo2015}, and the resulting 
luminosity functions are shown with blue dot-dashed and green long-dashed lines,
respectively, in the right panel of figure~\ref{fig:WIDE_z4QSO_XLF}. Converted
luminosity functions with and without the scatter are shown with thick and thin
 lines, respectively. The resulting luminosity function based on \citet{Masters2012} broadly overlaps with
that of the X-ray luminosity function, and they are apparently consistent with each other.
However, it should be noted that their quasar sample is selected based on similar
criteria to ours, employing blue UV color and stellar morphology.
Since we expect a significant contribution by obscured and/or fainter
AGN at least below $\log L_{\rm 2-10~keV}({\rm erg} \ {\rm s}^{-1})\sim43$,
this would imply an over estimation of their luminosity function at the faint-end.
Similarly, if we use the best-fit luminosity function in \citet{Giallongo2015},
the converted luminosity function significantly over-predict the number density,
especially below the knee of the luminosity function.

\section{Summary}

We construct a statistical sample of 1,666 $z=4$ quasars based on
339.8 deg$^{2}$ of $g$, $r$, $i$, $z$, $y$ band imaging data 
from the S16A-Wide2 release of the HSC-SSP program.
The $z=4$ quasar candidates are selected by their
stellar morphology and $g$-band dropout selection criteria.
Our selection is optimized for
selecting stellar objects with as low contamination from 
extended galaxies as possible. The selection is effective
down to $i=24$ mag. Our $g$-dropout color selection
criteria is based on the color distributions of SDSS 
quasars at $z=3.5-4.0$ and Galactic stars, which are
one of the most numerous contaminants to the selection.

The number counts of the selected $z=4$ quasars show
a monotonic increase in the covered magnitude range, $i=20.0-24.0$,
once we correct for the effect of contamination to the
selection. The survey selection efficiency and effective
survey area are evaluated with libraries of quasar photometric models
and a noise model of the HSC stacked images.
We evaluate the contamination by Galactic stars
and compact galaxies that meet the g-dropout selection
criteria. The results indicate that the contamination rate
can be more than 50\% in the magnitude range fainter than
$i>23.5$ mag, and the apparent excess seen in the magnitude range 
of the raw number counts
can be explained with this contamination.

Most of the $z=4$ quasar candidates do not have 
spectroscopic redshift information, and we estimate
photometric redshifts using a Bayesian method based on
our library of quasar models with templates .
The distribution of the resulting
photometric redshifts implies the sample covers the redshift
range between 3.6 to 4.3 with mean redshift of 3.9.
The distribution of the 
UV absolute magnitudes, $M_{\rm 1450}$, estimated from
the broad-band photometry, covers an absolute magnitude
range down to $M_{\rm 1450}\sim-22$ mag, which is more than
2 magnitudes deeper than the SDSS sample. 
Thanks to the large sample,
we determine the faint-end of the $z=4$ quasar
luminosity function with unprecedented statistics.
In order to extend our luminosity coverage to the bright-end,
we include the uniform $z=4$ quasar sample from the SDSS
DR7, providing a combined coverage of
$M_{\rm 1450}=-22$ to $-29$ mag.

The luminosity function is well described by
a double power-law model with a clear break around
$M_{\rm 1450}\sim-25$.
The bright-end slope, $\beta=-3.11\pm0.07$ is consistent with 
those derived at $z=2.2-3.5$ and at $z=5$. The faint-end
slope, $\alpha=-1.30\pm0.05$, is consistent with that at
$z=2.2-3.5$ obtained from BOSS, but flatter than that derived
at $z=3.2$ with deeper coverage from the COSMOS survey.
The overall shape of the $z=4$ luminosity function does
not support the higher break luminosities and
steeper faint-end slopes at higher redshifts suggested
at $z=5$ \citep{McGreer2013}. 

We convert the $M_{\rm 1450}$ luminosity function 
to the hard X-ray 2--10~keV luminosity function
using the relation between $l_{\rm 2500{\AA}}$ and 
$l_{\rm 2keV}$ \citep{Steffen2006}. Once we consider
the scatter of the relation, the number density of
UV selected quasars matches well with that of the
X-ray selected AGNs above the knee of the luminosity
function. Below the knee of the luminosity function, 
the UV-selected quasars show a deficiency compared
to the hard X-ray luminosity function. The deficiency
can be explained by obscured AGNs.

\bigskip
\begin{ack}
We thank Dr. Akio K. Inoue for providing us with the
updated Monte Carlo model of the IGM absorption.

The Hyper Suprime-Cam (HSC) collaboration includes 
the astronomical communities of Japan and Taiwan, and 
Princeton University. The HSC instrumentation and 
software were developed by the National Astronomical 
Observatory of Japan (NAOJ), the Kavli Institute for 
the Physics and Mathematics of the Universe (Kavli IPMU), 
the University of Tokyo, the High Energy Accelerator 
Research Organization (KEK), the Academia Sinica 
Institute for Astronomy and Astrophysics in Taiwan (ASIAA), 
and Princeton University. Funding was contributed by 
the FIRST program from Japanese Cabinet Office, the 
Ministry of Education, Culture, Sports, Science and 
Technology (MEXT), the Japan Society for the 
Promotion of Science (JSPS), 
Japan Science and Technology Agency (JST), 
the Toray Science Foundation, NAOJ, Kavli IPMU, KEK, 
ASIAA, and Princeton University. 

This paper makes use of software developed for the 
Large Synoptic Survey Telescope. We thank the LSST 
Project for making their code available as free 
software at \texttt{http://dm.lsst.org}.

The Pan-STARRS1 Surveys (PS1) have been made possible 
through contributions of the Institute for Astronomy, 
the University of Hawaii, the Pan-STARRS Project Office, 
the Max-Planck Society and its participating institutes, 
the Max Planck Institute for Astronomy, Heidelberg and 
the Max Planck Institute for Extraterrestrial Physics, 
Garching, The Johns Hopkins University, Durham University, 
the University of Edinburgh, Queen’s University Belfast, 
the Harvard-Smithsonian Center for Astrophysics, 
the Las Cumbres Observatory Global Telescope Network 
Incorporated, the National Central University of Taiwan, 
the Space Telescope Science Institute, the National Aeronautics 
and Space Administration under Grant No. NNX08AR22G issued 
through the Planetary Science Division of the NASA Science 
Mission Directorate, the National Science Foundation under 
Grant No. AST-1238877, the University of Maryland, and 
Eotvos Lorand University (ELTE) and the Los Alamos National Laboratory.

Based in part on data collected at the Subaru Telescope 
and retrieved from the HSC data archive system, which is 
operated by Subaru Telescope and Astronomy Data Center, 
National Astronomical Observatory of Japan.

Funding for the Sloan Digital Sky Survey IV has been provided by
the Alfred P. Sloan Foundation, the U.S. Department of Energy Office of
Science, and the Participating Institutions. SDSS-IV acknowledges
support and resources from the Center for High-Performance Computing at
the University of Utah. The SDSS web site is \texttt{www.sdss.org}.

SDSS-IV is managed by the Astrophysical Research Consortium for the 
Participating Institutions of the SDSS Collaboration including the 
Brazilian Participation Group, the Carnegie Institution for Science, 
Carnegie Mellon University, the Chilean Participation Group, the 
French Participation Group, Harvard-Smithsonian Center for Astrophysics, 
Instituto de Astrof\'isica de Canarias, The Johns Hopkins University, 
Kavli Institute for the Physics and Mathematics of the Universe (IPMU) / 
University of Tokyo, Lawrence Berkeley National Laboratory, 
Leibniz Institut f\"ur Astrophysik Potsdam (AIP),  
Max-Planck-Institut f\"ur Astronomie (MPIA Heidelberg), 
Max-Planck-Institut f\"ur Astrophysik (MPA Garching), 
Max-Planck-Institut f\"ur Extraterrestrische Physik (MPE), 
National Astronomical Observatories of China, New Mexico State University, 
New York University, University of Notre Dame, 
Observat\'ario Nacional / MCTI, The Ohio State University, 
Pennsylvania State University, Shanghai Astronomical Observatory, 
United Kingdom Participation Group,
Universidad Nacional Aut\'onoma de M\'exico, University of Arizona, 
University of Colorado Boulder, University of Oxford, University of Portsmouth, 
University of Utah, University of Virginia, University of Washington, University of Wisconsin, 
Vanderbilt University, and Yale University.
\end{ack}

\end{document}